\documentclass{aa}  
\usepackage{graphicx}
\usepackage{threeparttable}
\usepackage[leftcaption]{sidecap}
\sidecaptionvpos{figure}{b}
\usepackage{txfonts}
\usepackage{hyperref}
\hypersetup{
     colorlinks   = true,
     citecolor    = blue,
     linkcolor    = black,   
     urlcolor     = blue,
}

\begin{document} 

   \title{Study of the $\sim 50$ kpc circumgalactic environment around the \\ merger system J2057-0030 at $\MakeLowercase{z}$ $\sim$ 4.6 using ALMA}

    \titlerunning{Study of the $\sim 50$ kpc circumgalactic environment around the \\ merger system J2057-0030 at $\MakeLowercase{z}$ $\sim$ 4.6}
    
   \author{M. Fuentealba-Fuentes\inst{1,2} 
   \and P. Lira\inst{1,3}
   \and T. Díaz-Santos\inst{4,5}
   \and B. Trakhtenbrot\inst{6}
   \and H. Netzer\inst{6} 
   \and L. Videla\inst{7}
    }

   \institute{Departamento de Astronomía, Universidad de Chile, Camino el Observatorio 1515, Las Condes, Santiago, Casilla 36-D, Chile.
    \and
        International Centre for Radio Astronomy Research (ICRAR), The University of Western Australia, 35 Stirling Highway, Crawley, WA 6009, Australia.
    \and
        Millennium Nucleus on Transversal Research and Technology to Explore Supermassive Black Holes (TITANS), Camino del Sur 495, depto 504, 4130654 San Pedro de la Paz, Concepcion, Chile.
    \and
        Institute of Astrophysics, Foundation for Research and Technology–Hellas (FORTH), Heraklion, GR-70013, Greece.
    \and 
        School of Sciences, European University Cyprus, Diogenes street, Engomi, 1516 Nicosia, Cyprus.
    \and
        School of Physics and Astronomy, Tel Aviv University, Tel Aviv 69978, Israel.
    \and
        Joint ALMA Observatory, Alonso de Córdova 3107, Vitacura, Santiago 763-0355, Chile.
            }
   
\abstract  
    {We present ALMA band-7 observations of J2057-0030, a multi-component merger system at $z$ $\sim$ 4.68 spanning at least 50 kpc in size, using the [C$\textsc{ii}$] $\lambda$157.74 $\mu$m line and underlying far-infrared (FIR) continuum. We find two main components, the quasar (QSO) and a dusty star-forming galaxy (DSFG), both detected in [C$\textsc{ii}$] and continuum emission as well as multiple neighboring clumps detected only in [C$\textsc{ii}$]. Three of these clumps form a (tidal) tail that extends from the QSO in a straight direction to the west, covering a projected distance of $\sim$ 10 kpc. This perturbed morphology, added to a spatial distance of $\sim$ 20 kpc and a velocity offset of  $\Delta{v}$ = 68 km s$^{-1}$ between the QSO and the DSFG, strongly supports a merging scenario. By fitting a spectral energy distribution model to the continuum data, we estimate star formation rates of $\approx$ 402 $M_{\odot}$ yr$^{-1}$ for the QSO host and $\approx$ 244 $M_{\odot}$ yr$^{-1}$ for the DSFG, which locate them on or close to the main sequence of star-forming galaxies. The J2057-0030 QSO was selected for being one of the brightest unobscured quasars at its redshift while presenting a rather modest star formation rate. Based on a commonly accepted paradigm regarding the formation of quasars, this result is expected for a quasar that has already passed an obscured phase of rapid star formation during a major merger. However, we see that the merger event in this system is far from being finished, and it is rather likely somewhere between the first pericenter and subsequent close passages. This is presumably another case of a high-$z$ quasar residing in a high-density environment with a companion obscured galaxy.}


   \maketitle

\section{Introduction}\label{Chapter1}

The mass assembly of galaxies in early epochs of the Universe is expected to be a process driven by a combination of (at least) two main mechanisms: the accretion of cold gas from the intergalactic and circumgalactic medium (IGM and CGM) and the interaction between gas-rich galaxies, that is, major mergers \citep[e.g.,][]{Keres_2005, Dekel_2009, Hopkins_2006a}. These mechanisms are expected to support the supermassive black hole (SMBH) growth, during a phase of active galactic nuclei (AGN), and to simultaneously enhance the star formation (SF) of the host galaxy. This tight but complex connection between the SMBH and its galaxy host is supported by observational evidence, including the well-studied relations of the black hole mass (M$_\mathrm{BH}$) with the stellar velocity dispersion ($\sigma$) \citep[e.g.,][]{Ferrarese_2000, Gebhardt_2000, Gultekin_2009}, with the bulge luminosity \citep{Kormendy_1995}, and bulge mass \citep[e.g.,][]{Magorrian_1998, Haring_2004}. 

To test the interplay between mergers and inflows observationally, we need to focus on the early epochs of galaxy assembly, where the coevolution between SMBHs and their hosts should be taking place. However, obtaining direct observations of both mechanisms at high-$z$ is challenging, mainly because of the diffuse nature of the cold gas. Observations are also limited by the spatial resolution needed to resolve the structures and by the obstacle of differentiating between the emission of the SMBH and the galaxy host when the AGN emission dominates the optical-NIR spectral regime, diluting all properties of the host. Given these limitations, the best approach is to look into the far-infrared (FIR) regime where the dust grains heated by young stars dominate the continuum and when the host kinematics can be accessed by the observations of interstellar emission lines. 

Particularly, the [C$\textsc{ii}$] $\lambda$157.74 $\mu$m line arises from nearly every phase of the interstellar medium (ISM) due to its low ionization potential (11.26 eV) \citep{Lagache_2018} including the neutral, molecular, and ionized gas associated with dense photodissociation regions (PDRs) \citep{Hollenbach_1997, Wolfire_2003}. Additionally, the [C$\textsc{ii}$] line is the brightest FIR emission line \citep{Stacey_1991}, representing one of the most efficient coolants of the ISM. The measured transition frequency is 1900.537 GHz, corresponding to a transition wavelength of 157.74 $\mu$m and making the [C$\textsc{ii}$] line observable for 2 $\lesssim$ $z$ $\lesssim$ 8.5 using  observations from the Atacama Large Millimeter/submillimeter Array (ALMA).

Since the advent of ALMA, the evidence for gas accretion from the CGM has grown, with inflows detected up to z $\sim$ 1 \citep{Bouche_2016, Martin_2019, Zabl_2019}. Similarly, several works have reported interacting galaxies at high-$z$, where the presence of companion galaxies increases in high-density environments \citep[e.g.,][]{Trakhtenbrot_2017, DiazSantos_2018, Jones_2019, prochaska_2019, Ginolfi_2020, Nguyen_2020, Romano_2021, banerji_2021}. These mergers are identified based on their perturbed morphologies with extended distributions of gas and stars as seen in the formation of tidal tails, along with a small spatial projected distance ($\Delta{d}$ $<$ 50 kpc) and a small velocity offset between the galaxies ($\Delta{v}$  $<$ 500 km s$^{-1}$). 

In this paper, we present a study of the circumgalactic environment of J2057-0030, an ongoing merger composed of a quasar (QSO) at $z$ $\sim$ 4.68 and a companion dusty star-forming galaxy (DSFG) at a projected distance of $\sim$ 20 kpc, as previously determined by \citet{Trakhtenbrot_2017} and \citet{Nguyen_2020}. This analysis is based on ALMA cycle 6 observations of the [C$\textsc{ii}$] line and underlying FIR continuum emission. 
This paper is structured as follows: In Section \ref{Chapter2} we describe the data and methods of data reduction and spectral measurements. In Section \ref{Chapter3} we present the [C$\textsc{ii}$] and the dust continuum properties, including estimations of the dynamical, dust, and gas masses of the sources. In Section \ref{Chapter5} we discuss the nature of the system as a major merger. 
Conclusions are summarized in Section \ref{Chapter6}. Throughout this work we assume a cosmological model with $H_0$ = 70 km s$^{-1}$ Mpc$^{-1}$, $\Omega_{M}$ = 0.3, and $\Omega_{\wedge}$ = 0.7, which gives an angular scale of 6.5 kpc/$''$ at $z$ = 4.6. For the star formation rates (SFRs) we use a Chabrier initial mass function (IMF) from \citet{Chabrier_2003}.

%--------------------------------------------------------------------

\section{ALMA observations and data reduction}\label{Chapter2}

\subsection{Campaign background}\label{cap:CampaignBackground}

J2057-0030 is part of a flux-limited sample of luminous quasars in a narrow redshift bin around z$\sim$4.8.  Since the most powerful quasars at high-$z$ are likely the progenitors of the most massive galaxies at $z \sim 0$, samples such as this help to anchor models and observations of the most massive BHs and host galaxies over a period of about 12.5 billion years in the history of the universe.  The sample is based on the 38 brightest unobscured quasars from the sixth data release \citep[DR6,][]{Adelman_McCarthy_2008} of the Sloan Digital Sky Survey \citep[SDSS,][]{York_2000} at $z$ $\sim$ 4.65 - 4.92. $\mathrm{Mg\textsc{ii}}$ $\lambda$2798 line and continuum flux at 3000Å ($F_{3000}$) are measured using VLT/SINFONI and Gemini-North/NIRI spectroscopy and used to obtain  black hole mass and bolometric luminosity, presented in \citet{Trakhtenbrot_2011}. 
Additional observations by the Herschel Spectral and Photometric Imaging Receiver (SPIRE) are presented in \citet{Mor_2012} and \citet{Netzer_2014}. Spitzer Infrared Array Camera (IRAC) photometry at 3.6 and 4.5 $\mu$m reported in \citet{Netzer_2014} are used as positional priors for the Herschel photometry. The sources detected by the Herschel/SPIRE instrument are divided into detected FIR-Bright and undetected FIR-faint sources. The first group is characterised by FIR luminosities of $L_{\mathrm{8-1000}}$ $\sim$ 8.5 $\times$ $10^{46}$ erg s$^{-1}$ %(2.2 $\times$ $10^{13}$ $L_{\odot}$) 
suggesting SFRs in the range $\sim$ 1000 - 4000 $M_{\odot}$ yr$^{-1}$. The $L_{\mathrm{8-1000}}$ of the undetected sources are estimated from a stacking analysis which resulted in a median SFR of about 400 $M_{\odot}$ yr$^{-1}$ \citep[][]{Netzer_2014} .

In \citet{Nguyen_2020} (N20 hereafter) ALMA Cycle 4 observations of twelve of these luminous quasars are reported including our system. J2057-0030 corresponded to a FIR-Faint source with luminosity $L_{\mathrm{8-1000}}$ $\sim$ $10^{46.83}$ erg s$^{-1}$, accretion rate of $L/L_{\mathrm{Edd}}$ $\sim$ 0.89, and $M_{\mathrm{BH}}$ $\sim$ $10^{9.23}$ $M_{\odot}$ as reported in \citet{Trakhtenbrot_2011}. 

Archival observations from Wide-field Infrared Survey Explorer \citep[WISE;][]{Wright_2010} and the Galaxy Evolution Explorer \citep[GALEX;][]{Martin_2005} are available for the system but the angular resolution is too low ($\sim$ 6$''$ and 4$''$, respectively) to attempt any derivation of the physical properties.

\subsection{ALMA observations}\label{cap:ALMAObservations}

J2057-0030 was previously observed with ALMA band-7 (275 – 373 GHz) during Cycle 4 (see N20 for further details). However, the rejection of a few channels at the edge of the spectral windows (SPWs) due to divergent flux values led to a small gap in the spectra. Unfortunately, the [C$\textsc{ii}$] emission of all sources associated with the J2057-0030 system lay in the middle of the gap, resulting in a significant loss of spectral information.

In this work, we present low and high spatial resolution observations for J2057-0030 obtained with ALMA band-7 during Cycle 6 (PI: Nguyen N., PID: 2018.1.01830.S). The 
low resolution data has a synthesized beam size of 0.42$''$ $\times$ 0.36$''$ at position angle (P.A.) = -35.69° and it was taken using the C43-4 configuration with an on-source time of $\sim$ 19 minutes. For the high resolution data, the C43-7 configuration was used 
during 2 execution blocks. The on-source time was 40 minutes per block, resulting in 
a synthesized beam size of 0.12$''$ $\times$ 0.1$''$ at P.A. = -52.96°. In this cycle, J2253+1608 was used for amplitude, band-pass, and pointing calibration, while the source J2101+0341 was used for phase calibration.

For both resolution setups, the spectral setup consisted of two sidebands separated by about 12 GHz, each constructed of one SPW. For the [C$\textsc{ii}$] line window, the FDM correlator mode was selected covering an effective bandwidth of 1875 MHz, which corresponds to $\sim$ 1650 km s$^{-1}$. This spectral range is sampled by 3840 channels with a frequency of 0.488 MHz or $\sim$ 0.437 km s$^{-1}$ per channel. For the continuum spectral window, the TDM correlator mode was used (same as N20), again covering an effective bandwidth of 1875 MHz ($\sim$ 1650 km s$^{-1}$), but sampled by 128 channels yielding a frequency of 15.625 MHz ($\sim$ 15 km s$^{-1}$).

\subsection{Data reduction}\label{cap:DataReduction}

Data reduction was performed using the CASA package version $5.6.1$ \citep{McMullin_2007}. All Cycle 6 line-free SPWs at different spatial resolutions were concatenated into a single measurement set (MS file) using the CONCAT command. This file was used to create the continuum images. For the [C$\textsc{ii}$] line cubes, the continuum was subtracted in each of the SPWs at different resolutions independently using UVCONTSUB. The cubes of different resolution setups were concatenated back into a single MS. 

\begin{figure*}
\centering
\includegraphics[scale=0.45]
{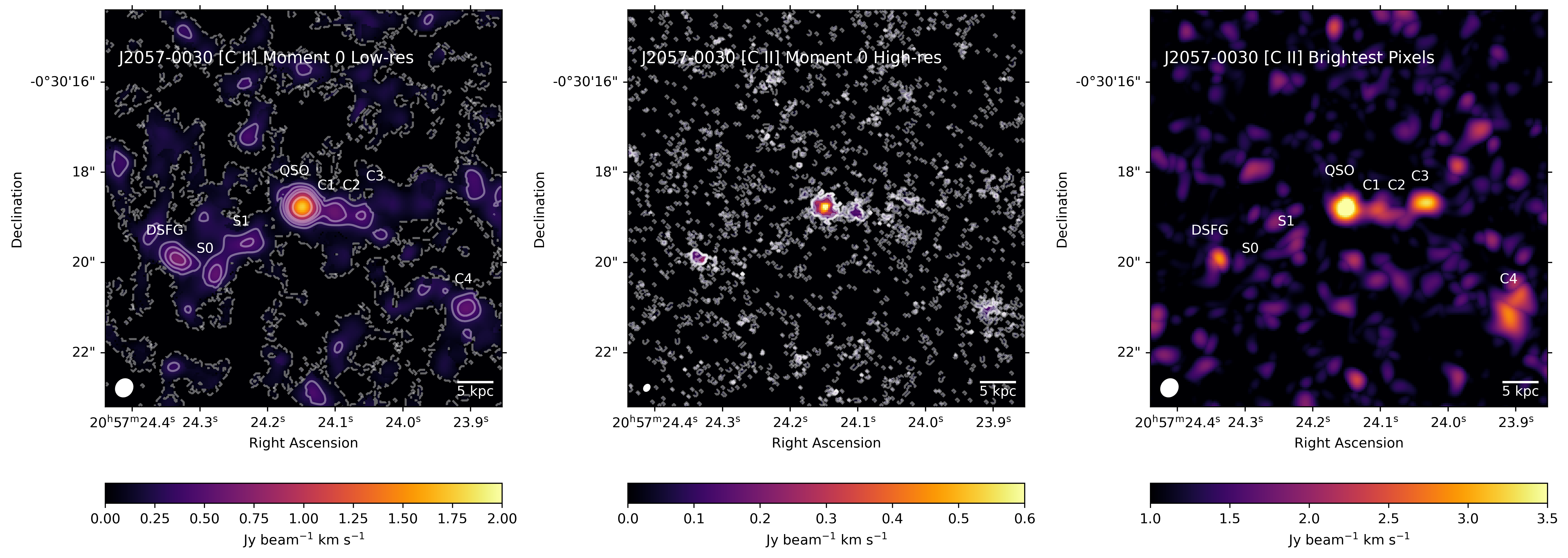}
\caption{Components of the system are presented: the QSO, the DSFG, and the clumps (C1, C2, C3, C4, S0, and S1). $\textbf{Left}$: The map is showing the integrated value of the [CII] spectrum (moment 0) for the lower resolution combination. The beam size is 0.41$''$ $\times$ 0.36 at P.A = -35.69°. Solid (dashed) white contours indicate the positive (negative) significance levels at [2, 3, 4, 5, 8] $\sigma$. $\textbf{Middle}$: Moment 0 for the higher resolution combination. The beam size is 0.16$''$ $\times$ 0.12$''$ at P.A = -35.69°. White solid (dashed) contours indicate the positive (negative) significance levels at [2, 3, 4, 5, 8] $\sigma$. $\textbf{Right}$: The map is showing the brightest spaxel value of the [C$\textsc{ii}$] spectrum for the lower resolution combination. C3 is only noticeable in one channel as it has a narrow FWHM $\sim$ 82 km s$^{-1}$. For all the panels, the synthesized beam is shown in the bottom-left corner as a white filled ellipse. North is up and east is to the left.}
\label{mom0_contours}
\end{figure*}

For the line cube, we averaged the flux distributed in 3840 channels by applying the TCLEAN command to channels of 54 km s$^{-1}$ width. We applied a "briggs" weighting with robust = 2.0. In order to explore the spatial distribution of the sources and also measure their spectral properties, we used the "restoringbeam" parameter with two different values which yielded two cubes with different spatial resolutions. 
The first beam size was set to $0.41'' \times 0.36''$ at a  P.A. = -35.69° which is closer to the low resolution beam size and appears to provide a convenient compromise between resolution and signal-to-noise (S/N). The other beam size was $0.16'' \times 0.12''$ at a P.A = -35.69°, allowing multiscale analysis of our sources (see Figure \ref{mom0_contours}).

\subsection{Source detection}

\begin{figure*}[!h]
\centering
\includegraphics[scale=0.4]{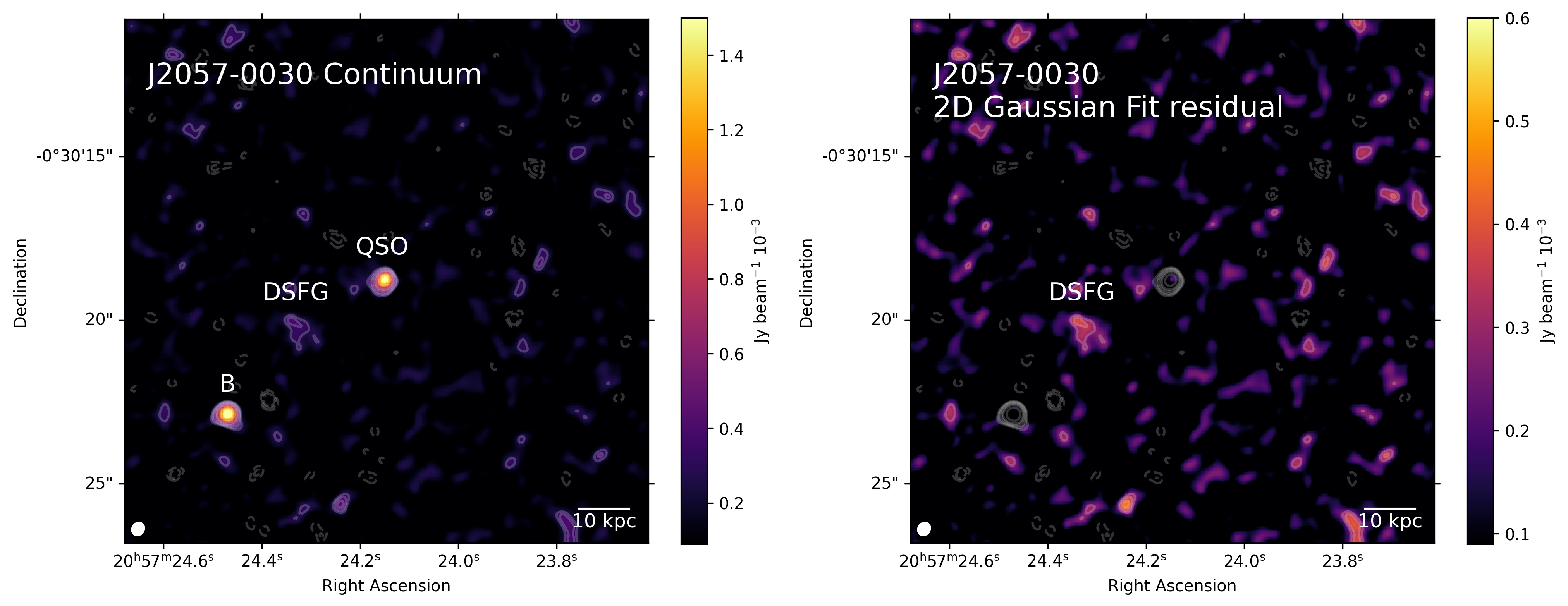}
\caption{Continuum emission images. $\textbf{Left}$: The original image. The QSO has significant emission as well as the source "B" which has no [C$\textsc{ii}$] emission and as \citet{Nguyen_2020} mentioned it seems to be a source only seen in projection. The DSFG shows faint continuum emission. $\textbf{Right}$: The QSO and the source "B" have been fitted by a CASA 2D Gaussian and have been subtracted from the original image. Both panels have the contours over-plotted as solid (dashed) white contours indicating the positive (negative) significance levels at [2, 3, 4, 5, 8] $\sigma$. The synthesized beam is shown at the bottom-left as a white filled ellipse. North is up and east is to the left.}
\label{Cont}
\end{figure*}

\begin{SCfigure*}
\centering
\includegraphics[scale=0.365
]{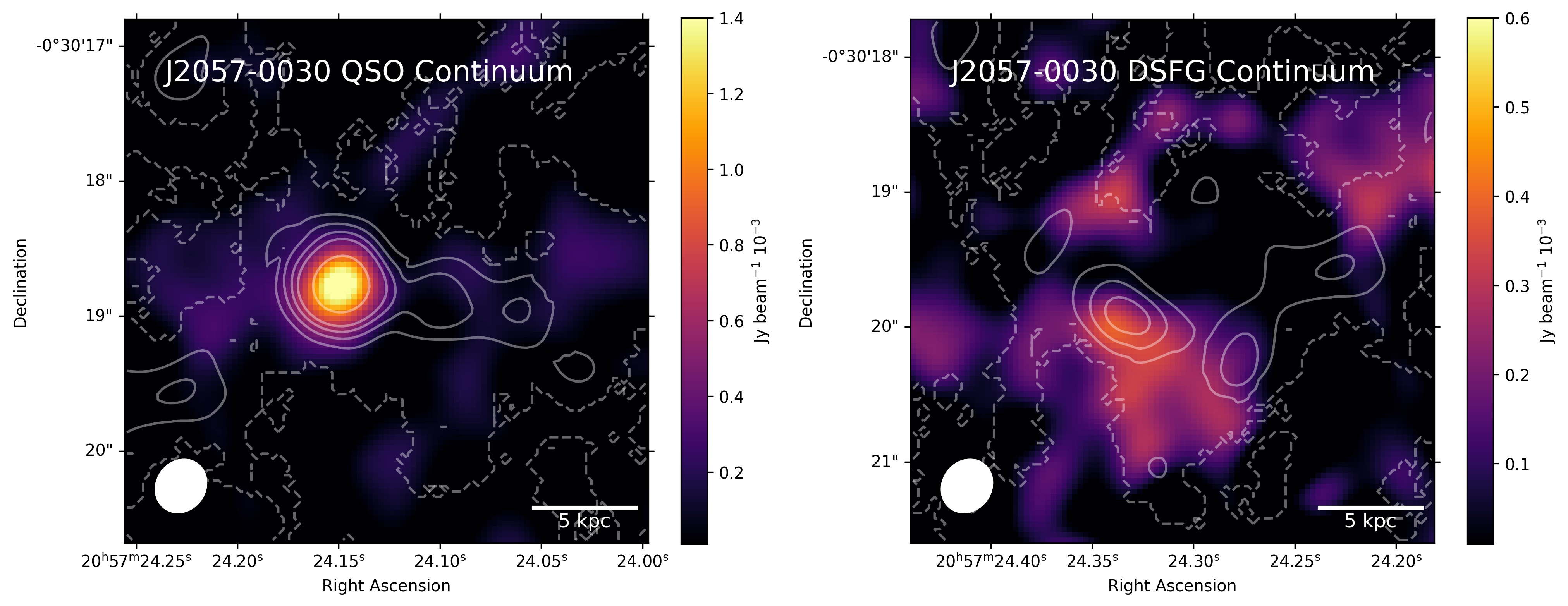}
\caption{Continuum emission with [C$\textsc{ii}$] contours over-plotted as solid (dashed) white contours indicating the positive (negative) significance levels at [2, 3, 4, 5, 8] $\sigma$. $\textbf{Left}$: QSO continuum emission image plus the nondetected clumps C1, C2, and C3. $\textbf{Right}$: DSFG continuum image with the neighboring clumps, S0 and S1. The synthesized beam is shown at the bottom-left corner as a white filled ellipse. North is up and east is to the left.}
\label{Cont_qso_DSFG}
\end{SCfigure*}

%-------------------------------------------------------------

\subsubsection{[C$\normalfont\textsc{ii}$] emission}

Figure \ref{mom0_contours} left panel shows the integrated flux of the spectrum (moment 0) of the [C$\textsc{ii}$] line. The QSO is detected as well as the DSFG (previously identified as a submillimeter galaxy (SMG) in N20\footnote{We label the companion as a dusty star-forming galaxy (DSFG) instead of a submillimeter galaxy (SMG) as the continuum flux density of this source is 1.4 mJy, half of the flux expected for an SMG S$_{850}$ $>$ 3-5 mJy.}), and a tail-like extended emission, which in contrast with N20, breaks into three distinct clumps projected in an almost straight direction from the QSO westward. They are labeled as C1, C2, and C3. Their [C$\textsc{ii}$] integrated fluxes have S/N ratios\footnote{The signal-to-noise (S/N) is calculated as the integrated value of the [C$\textsc{ii}$] divided by the squared sum of the noise over the integrated range of the spectrum of $\pm$ 3$\sigma$ $\left(F/\sqrt{\sum_i{n_i^2}}\right)$.} of $\sim$8, 6, and 5, respectively. C3 corresponds to a narrow feature with a full width at half maximum (FWHM) $\sim$ 82 km s$^{-1}$. Considering that the width of the cube channels is 50 km s$^{-1}$, the [C$\textsc{ii}$] line corresponds approximately to one channel and it is identified due to its particularly strong brightness in a map of the peak flux of every spaxel (see Figure \ref{mom0_contours} right panel). We explored whether a different spectral binning could improve the study of C3 by using a channel width of $\sim$ 20, 25, and 30 km s$^{-1}$. In the three cases, the spectra of C3 and the Gaussian fit agree well with the results obtained using a width of $\sim$ 50 km s$^{-1}$ and the integrated fluxes do not change significantly.
We detect a fourth clump, labeled as C4, at $\sim$ 8$\sigma$ in the southwest direction and it is the furthest clump from the QSO at a projected distance of $\sim$ 28 kpc.
In addition to our six components, we detect two fainter clumps to the west of the DSFG, which we refer to as S0 and S1 (see Figure \ref{mom0_contours}) with their [C$\textsc{ii}$] integrated fluxes detected at $\sim$ 6$\sigma$. Throughout this work, we discuss the possibility of S0 and S1 being real sources comprising a "tidal bridge."

\subsubsection{Continuum emission}

We detect the QSO host and the DSFG in continuum emission. The clumps close to the QSO remain undetected. These results are in agreement with N20 which previously reported no continuum in the regions associated with the tidal tail. The clumps close to the DSFG (S0 and S1) are also not detected in dust continuum.

An extra source "B" is detected, previously reported by N20 as a source only seen in projection as it is bright in continuum but lacking detection in [C$\textsc{ii}$]. Besides ALMA, "B" is not detected in any other ancillary data catalog. If "B" is another component physically associated with the system, its FIR luminosity would correspond to $\log L_{\mathrm{FIR}}$/$L_{\odot}$ $\sim$ 12.7 with a projected distance to the QSO of $\sim$ 6.3$''$ (42 kpc). 

A rough estimation of the [C$\textsc{ii}$] emission for "B" gives an upper limit for the $L_{\mathrm{C\textsc{ii}}}$/$L_{\mathrm{8-1000}}$ ratio (see Section \ref{cap:Lratio}), corresponding to an extreme [C$\textsc{ii}$] deficit. Alternatively, the non-detection of [C$\textsc{ii}$] could represent a large velocity offset with respect to the QSO rest frame ($\Delta{v}$ $>$ 700 km s$^{-1}$), that is therefore not included in our SPW of width of $\sim$ 1500 km s$^{-1}$. We stress that this scenario would be in agreement with having a large spatial separation with respect to the QSO. As we are not able to conclude if "B" is part of the system, we only report on its continuum properties but do not include it in the discussion of the J2057-0030 merger.

To improve the visualization of the fainter sources, such as the DSFG and the clumps, against the background, we fit a 2D Gaussian to the QSO and source "B." Then we characterise the sizes of their continuum emitting regions in order to subtract the QSO and "B" emission from the continuum image. This image is not used for flux measurements. The result is seen in Figure \ref{Cont} right panel, where the limits of the colorbar had been adjusted in order to highlight the emission of the DSFG, which shows an extended morphology. In fact, in N20 the DSFG has two components labeled as NE and SW according to their position, which is in good agreement with the strongest emission found in our case, which are aligned in that direction. Figure \ref{Cont_qso_DSFG} also shows more low surface brightness emission that extends to the south of the DSFG. However, it is not possible to confirm its detection with the available sensitivity as the emission does not truly stand out from the background.
Because of the faint emission at the south of the DSFG, the 2D Gaussian fit of the continuum emission yields a source that is larger than deduced from [C$\textsc{ii}$], where the line emission is more concentrated and the extra [C$\textsc{ii}$] surrounding it is considered to be comprised of two different clumps that are fitted independently. It is important to remark that the location of the [C$\textsc{ii}$] clumps poorly match the distribution of the continuum emission, as only the S0 clump might partially coincide with one of the extended continuum regions.
This behavior of the DSFG with a larger extension of the FIR continuum (which traces dust heated by UV emission from star-forming regions), than the [C$\textsc{ii}$] emission is the opposite of what has been seen in the literature for sources at high-$z$ ($z >$ 4) \citep[e.g.,][although they mostly focus on less FIR-luminous main sequence SFGs]{Fujimoto_2019, Carniani_2020}, where the [C$\textsc{ii}$] regions are more extended than the FIR continuum. A more complete determination of the morphology of the DSFG in [C$\textsc{ii}$] and dust continuum will require deeper observations.

\subsection{Continuum and spectral measurements}

We measure the FIR continuum fluxes of the sources with two approaches depending on the nature of each source. In the first case, we calculate the sizes of the continuum regions of the QSO and the DSFG by fitting a CASA 2D Gaussian, which is characterised by a peak flux, semi-major and semi-minor axes, and a position angle. The fluxes of the QSO and its companion are measured by integrating over the beam-convolved sizes given by the 2D Gaussians. For the remaining components, CASA cannot fit a spatial 2D Gaussian as the emission of the sources is too weak. For those components, the fluxes are obtained directly from the continuum images using elliptical apertures based on the [C$\textsc{ii}$] extension of the sources. 

Using the [C$\textsc{ii}$] line cubes we calculate the zero, first and second moments corresponding to the integrated value of the spectrum, the velocity shifts and velocity dispersion fields, respectively. To create them, we use
LMFit to fit a Gaussian profile along the velocity axis of every spaxel in the cube using three free parameters: the amplitude ($I_0$), the standard deviation ($\sigma$) and the center of the Gaussian ($x_0$). Then, we apply moment equations directly over the spectral ranges enclosed within $\pm3\sigma$, with $\sigma$ given by the best fit, as this range is expected to include $\sim$ 99.73 $\%$ of the total line flux in the spaxel.

We use the moment zero map to determine the [C$\textsc{ii}$] sizes by fitting a CASA 2D Gaussian to each component of the system (as explained above), except for C3 which is our narrowest feature. In this case we use an elliptical aperture based in our visual estimation of its [C$\textsc{ii}$] extension instead. We define elliptical apertures to extract 1D spectra of the [C$\textsc{ii}$] sources using a radius of 2$\sigma$ based on the 2D Gaussian estimation of their beam-convolved sizes, representing $\sim$ 86.47$\%$ of the total flux of the sources if their spatial flux distributions is well described by a 2D Gaussian. We do not correct these values to account for the percentage of  missing light and therefore the fluxes of all the components should be larger. The sizes of the apertures are chosen in order to avoid the spatial overlapping of the main sources and clumps. To estimate the line fluxes we fit a Gaussian profile to the [C$\textsc{ii}$] spectrum using LMFit and then integrated over the line within $\pm$ 3$\sigma$. [C$\textsc{ii}$] and continuum beam-deconvolved sizes of the sources are reported in Table \ref{fluxes} with the uncertainties estimated using Monte Carlo simulations. 

\section{[{C$\normalfont\textsc{ii}$] line and dust continuum properties}} \label{Chapter3}

In this section and hereafter, the fluxes in the combined [C$\textsc{ii}$] and continuum data are scaled, to correct for the differences in peak and integrated fluxes when comparing the combined high and low resolution data with the individual data sets at different resolutions. See Appendix \ref{cap:Calibrations} for further details on the calibration issues and the scaling factor.

\subsection{[{C$\normalfont\textsc{ii}]$} line properties}\label{cap:1DSpectra}

\begin{figure}
\centering
\includegraphics[scale=0.32]{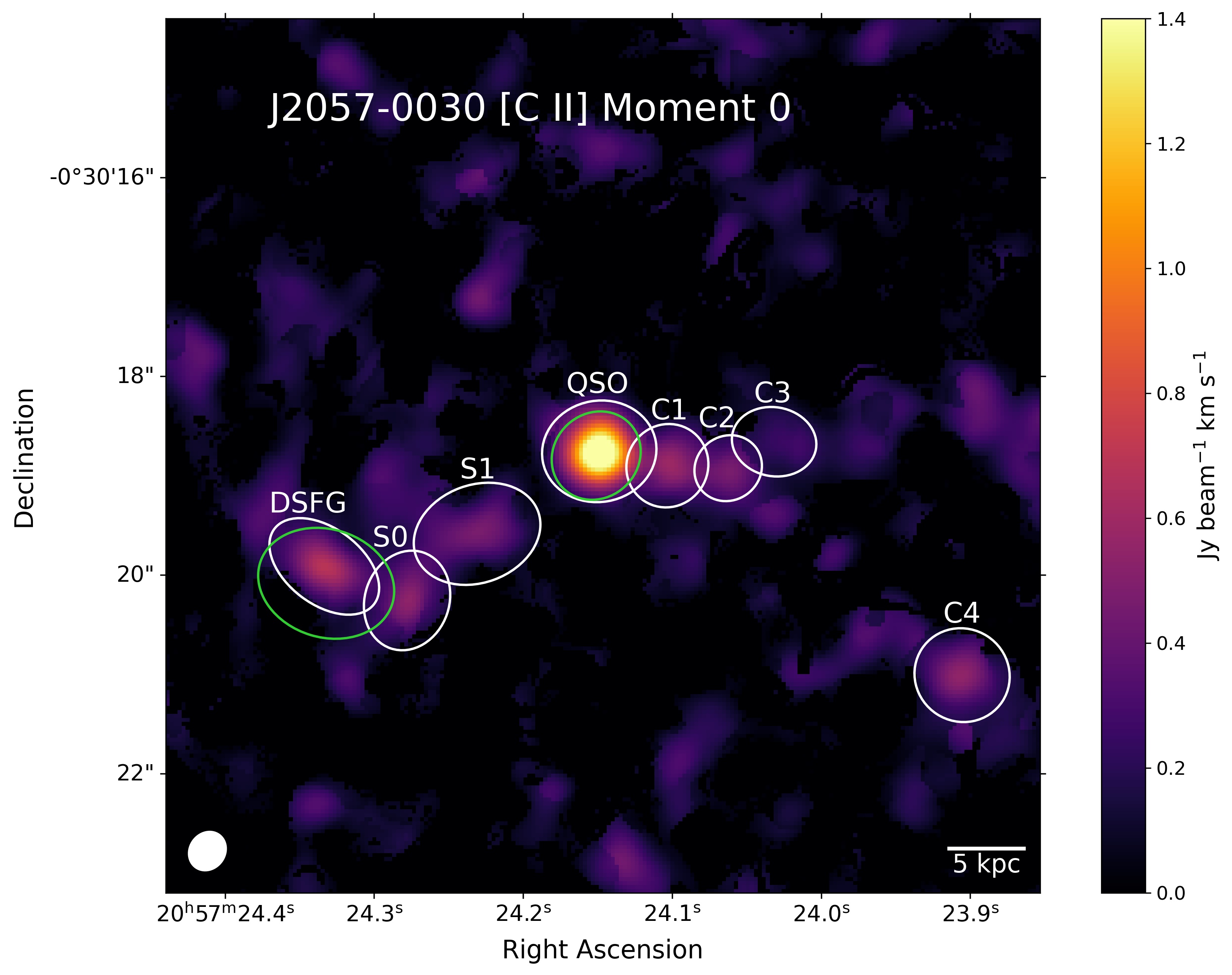}
\caption{Apertures used for flux measurements. The map shows the integrated value of the [C$\textsc{ii}$] spectrum (moment 0) for the lower resolution combination. The white ellipses represent: The QSO, the DSFG, and the clumps (C1, C2, C3, C4, S0, and S1). The white apertures are used to extract the [C$\textsc{ii}$] 1D spectra and to measure the continuum of the clumps. The continuum of the QSO and DSFG are estimated using the green ellipses representing the 2D Gaussian fit of the continuum. The synthesized beam is shown in the bottom-left as a white filled ellipse. North is up and east is to the left.}
\label{apertures}
\end{figure}

\begin{figure*}
\centering
\includegraphics[scale=0.53]{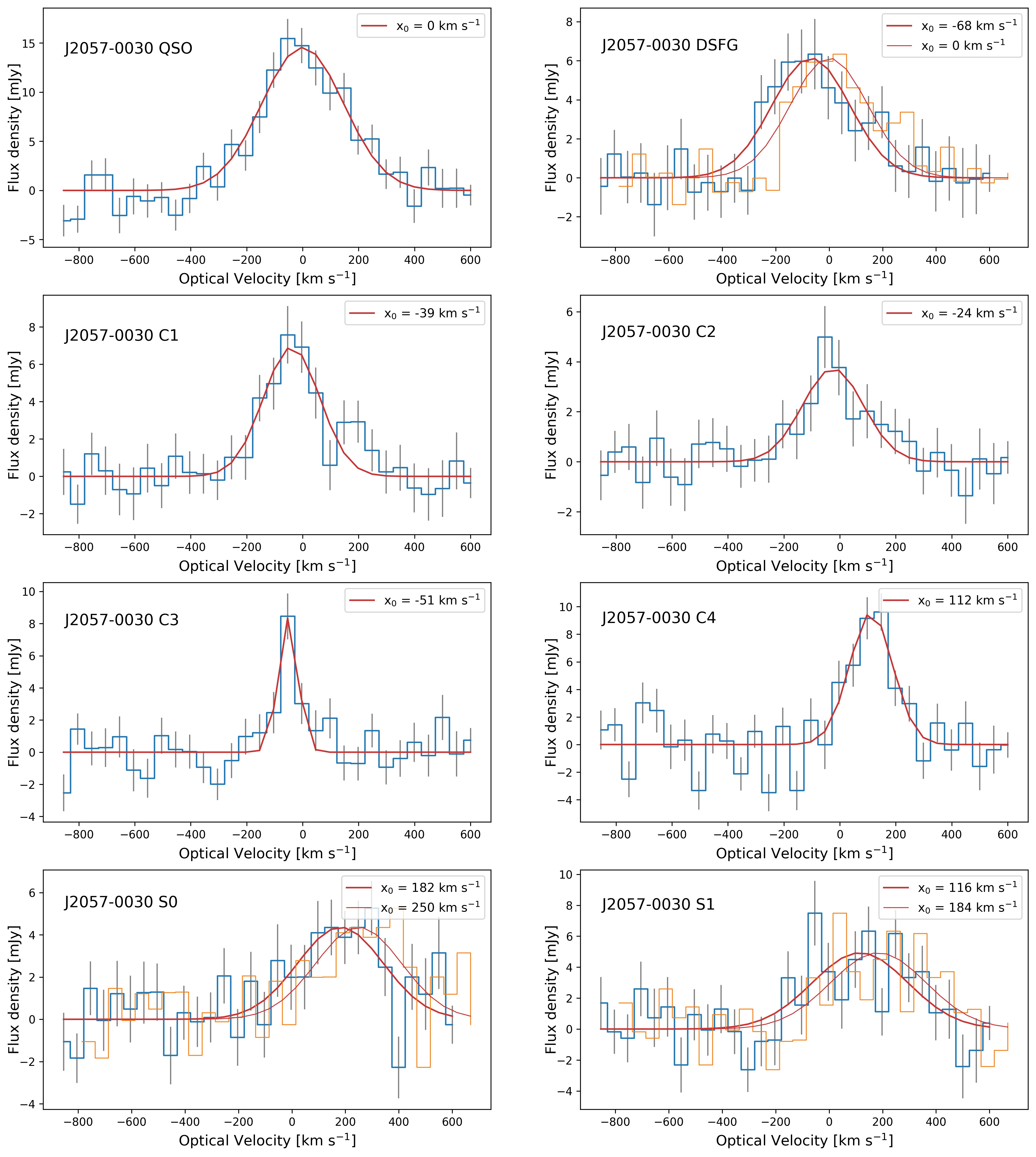}
\caption{Spectra of the [C$\textsc{ii}$] $\lambda$157.74 $\mu$m emission line for all the components of the system. A 1D Gaussian fit is shown as a solid red line with its respective center x$_0$ shown in the right corner. The orange line represents the spectra of the sources centered at the $x_0$ of the DSFG. The rms is plotted as gray bars.}
\label{spectra}
\end{figure*}

% Plot, Apertures 
We extract the 1D spectra of our components using elliptical apertures based on the beam-convolved sizes from the 2D Gaussian fit of the [C$\textsc{ii}$] emission, which are shown in Figure \ref{apertures} in white for the 6 main components analyzed in this work: the QSO, the DSFG and the clumps C1, C2, C3, C4, S0, and S1. The green ellipses displayed in Figure \ref{apertures} represent the 2D Gaussian fit of the continuum for the two main sources. The QSO and the DSFG are marginally resolved in [C$\textsc{ii}$], while the rest of the clumps have sizes comparable to the beam.
Figure \ref{spectra} shows the [C$\textsc{ii}$] spectra of the sources as step plots, while the respective Gaussian fits are shown as a solid red line. Velocities are relative to the peak intensity of the QSO fit. For the QSO, the [C$\textsc{ii}$] line covers a wide range of frequencies with a FWHM $\sim$ 345 km s$^{-1}$ and a peak of $\sim$ 11 mJy, in good agreement with N20. The spectrum is not completely symmetric showing slightly more flux at higher velocities ($\sim$ 200 km s$^{-1}$). However, we notice that a single Gaussian fit still works better than fitting two components. 

The DSFG spectra indicates a velocity shift with respect to the QSO’s systemic redshift of about $\Delta{v}$ = 68 km s$^{-1}$ ($\Delta{z}\approx$ 0.0015). This small velocity offset reinforces our idea of an interaction between the QSO and the DSFG as their velocity difference is comparable to (and in some cases smaller than) other shifts seen in ongoing mergers at high-$z$ \citep[e.g.,][]{Jones_2019, Ginolfi_2020, Romano_2021}.  
The emission of the DSFG consists of a peak of $\sim$ 5 mJy. 

The clumps C1, C2, and C3 show velocity offsets with respect to the QSO of $\Delta{v}$ = 39 km s$^{-1}$, 24 km s$^{-1}$,and 51 km s$^{-1}$, respectively, suggesting that all of them could be gravitationally bound to the system. Meanwhile C4, the spatially furthest clump from the QSO, shows the most extreme velocity offset with 112 $\pm$ 12 km s$^{-1}$. 
The Gaussian fit for C3 presents the poorest result as its emission is very narrow and likely subsampled, covering a few kilometers per second (one channel). S0 and S1 show velocity shifts with respect to the QSO’s systemic redshift of $\Delta{v}$ = 181 km s$^{-1}$ and 116 km s$^{-1}$, respectively. If we compare their velocities to the DSFG’s systemic redshift we obtain a velocity offset of $\Delta{v}$ = 249 km s$^{-1}$ and 184 km s$^{-1}$. 

%-----------------------------------------------------------------

\newpage
\begin{sidewaystable} \small
\begin{threeparttable}
\label{fluxes}
\caption{Spectral measurements.}
\begin{tabular}{c c c c c c c c c c c c c}
	\hline
	\hline
 
Comp. & Cont. Flux & $\nu$ & Cont. Size & S/N$_{\mathrm{Cont}}$ & F$_{[\mathrm{C\textsc{ii}}]}$ & $\mathrm{FWHM_{[C\textsc{ii}]}}$ & $\nu_0,_{[\mathrm{C\textsc{ii}}]}$ & [C$\textsc{ii}$] Size & $L_{[\mathrm{C\textsc{ii}}]}$ & $\Delta{d}$ & $\Delta{v}$ & S/N$_{[\mathrm{C\textsc{ii}}]}$ \\
 & [mJy] & [GHz] & [$''$] & & [Jy km s$^{-1}$] & [km s$^{-1}$] & [GHz] & [$''$] & [10$^9$ $L_{\odot}$] & [kpc] & [km s$^{-1}$] \\ \hline

QSO & 2.44 $\pm$ 0.24 & 346.52 & 0.76 $\pm$ 0.04 $\times$ 0.68 $\pm$ 0.03 & 10 & 3.79 $\pm$ 0.24 & 344.9 $\pm$ 49.6 & 334.45 & 1.14 $\pm$ 0.05 $\times$ 0.92 $\pm$ 0.05 & 2.59 $\pm$ 0.16 & ... & ... & 16\\
DSFG & 1.48 $\pm$ 0.33 & 346.52 & 1.48 $\pm$ 0.3 $\times$ 1.0 $\pm$ 0.2 &  4.5  & 1.58 $\pm$ 0.22 & 342.6 $\pm$ 68.8 & 334.53 & 1.28 $\pm$ 0.13 $ \times$ 0.34 $\pm$ 0.03 & 1.08 $\pm$ 0.15 & 19.4 & 68.1 & 7\\
C1$^{a,b}$ & 0.29 $\pm$ 0.22 & 346.52 &  $<$ 0.64 $\times$ 0.58 & 1 & 1.24 $\pm$ 0.16 & 238.2 $\pm$ 51.8 & 334.5 &  $<$ 0.64 $\times$ 0.58 & 0.85 $\pm$ 0.11 & 4.5 & 39.5 & 8\\
C2$^{a,b}$ & 0.23 $\pm$ 0.18 & 346.52 &  $<$ 0.82 $\times$ 0.76 & 1 & 0.72 $\pm$ 0.13 & 256.7 $\pm$ 70.2 & 334.48 &  $<$ 0.82 $\times$ 0.76 & 0.49 $\pm$ 0.09 & 8.5 & 24.5 & 6\\
C3$^{a,b}$ & 0.49 $\pm$ 0.21 & 346.52 &  $<$ 1.01 $\times$ 0.82 & 2 & 0.52 $\pm$ 0.1 & 81.8 $\pm$ 24.4 & 334.51 &  $<$ 1.01 $\times$ 0.82 & 0.35 $\pm$ 0.07 & 11.4 & 51.3 & 5\\
C4$^{b}$ & 0.38 $\pm$ 0.25 & 346.52 &  $<$ 0.88 $\times$ 0.74 &  1.5  & 1.3 $\pm$ 0.16 & 180.8 $\pm$ 58.2 & 334.33 & 0.88 $\pm$ 0.09 $\times$ 0.74 $\pm$ 0.07 & 0.89 $\pm$ 0.11 & 27.8 & 111.9 & 5\\
S0$^{b,c}$ & 0.52 $\pm$ 0.25 & 346.52 &  $<$ 0.92 $\times$ 0.68 & 2 & 1.24 $\pm$ 0.22 & 378.6 $\pm$ 148.0 & 334.34 & 0.92 $\pm$ 0.14 $\times$ 0.68 $\pm$ 0.1 & 0.85 $\pm$ 0.15 & 5.8 & 249.7 & 6\\
S1$^{b,c}$ & 0.04 $\pm$ 0.3 & 346.52 &  $<$ 1.32 $\times$ 0.88 & 0 & 1.57 $\pm$ 0.28 & 421.8 $\pm$ 182.8 & 334.42 & 1.32 $\pm$ 0.26 $\times$ 0.88 $\pm$ 0.18 & 1.07 $\pm$ 0.15 & 10.2 & 184.0 & 6\\ \hline

\end{tabular}
\vspace{0.25cm}
\begin{tablenotes}\small
\item \textbf{Note.} Columns: (1) Component; (2) Continuum fluxes (in mJy); (3) Continuum frequency (in GHz); (4) Beam deconvolved sizes of the continuum emitting region (in arcseconds); (5) [C$\textsc{ii}$] integrated fluxes (in Jy km s$^{-1}$); (7) Full width at half maximum (FWHM) of the Gaussian fit of the [C$\textsc{ii}$] 1D spectra (in km s$^{-1}$); (8) Observed central frequency of the [C$\textsc{ii}$] line (in GHz); (9) Beam deconvolved sizes of the [C$\textsc{ii}$] emitting region (in arcseconds); (10) [C$\textsc{ii}$] luminosities (in $L_{\odot}$); (11) Distances between the sources and the QSO host, calculated assuming the redshifts of the QSO hosts’ [C$\textsc{ii}$] emission lines (in kpc); (12) Velocity offsets with respect to the QSO host, calculated from the central frequencies of the [C$\textsc{ii}$] emission lines (in km s$^{-1}$); (13) Signal-to-noise ratio of the [C$\textsc{ii}$] emission. 
\item $^a$  [C$\textsc{ii}$] unresolved sources, sizes have upper limits only.
\item $^b$  Continuum unresolved sources, sizes have upper limits only.
\item $^c$ Distances in column (11) are measured relative to the DSFG. 
\end{tablenotes}
\end{threeparttable}
\end{sidewaystable}

\clearpage
\subsection{[C$\normalfont\textsc{ii}$] moments}\label{cap:Moments}

The left and middle panel of Figure \ref{mom0_contours} display the moment 0 map 
as seen in two different resolutions. Solid white
contours indicate the positive 
significance levels at [2, 3, 4, 5, 8]$\sigma$ of [C$\textsc{ii}$] flux. On the left, we show the map for the low resolution observations, where $\sigma_{[\mathrm{C\textsc{ii}]}}$ = 0.218 Jy km  s$^{-1}$ beam$^{-1}$. On the middle, we show the moment 0 map for the high resolution combination with $\sigma_{[\mathrm{C\textsc{ii}]}}$ = 0.046 Jy km  s$^{-1}$ beam$^{-1}$. Some complex structure can be seen toward the north and to the southwest of the QSO at lower brightness levels, implying that we could be starting to resolve the QSO host. The same occurs at the position of the DSFG. The rest of the components are seen as faint sources, not very distinguishable from the background (S/N $<$ 3).

\begin{SCfigure*}
\centering
\includegraphics[scale=0.34]{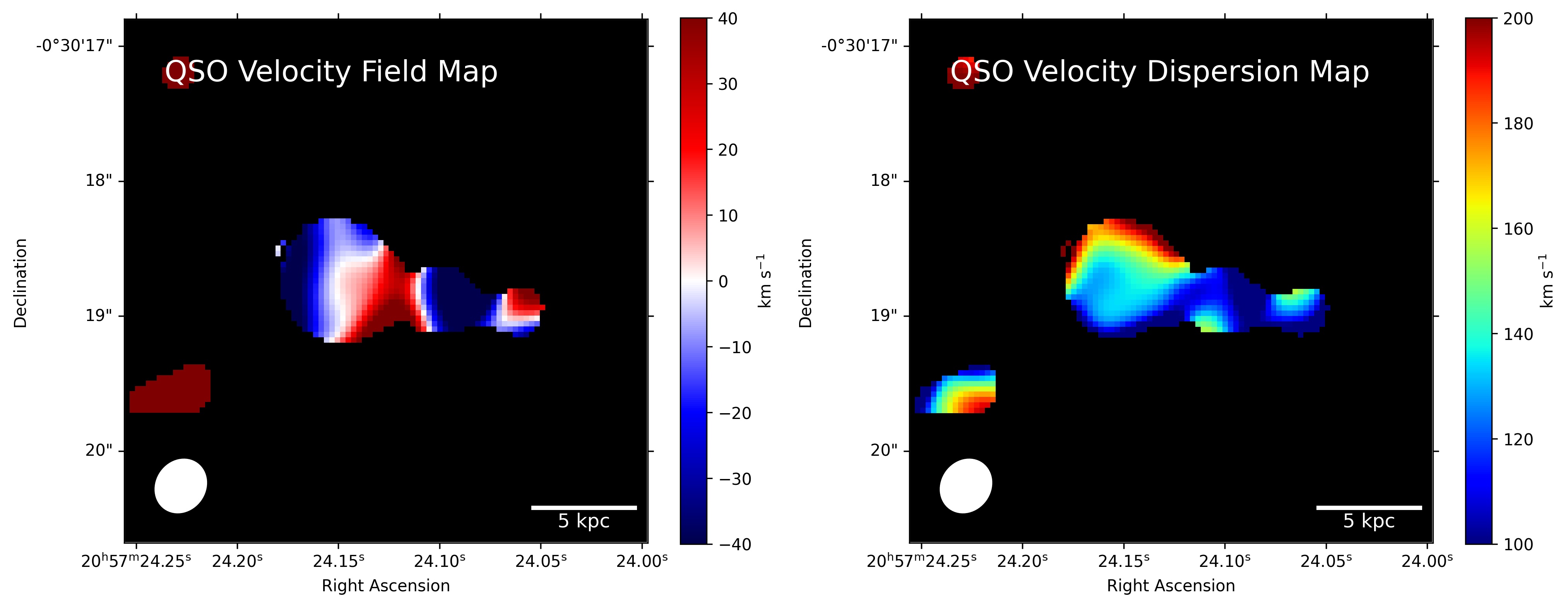}
\caption{Map of the velocity field (moment 1) and velocity dispersion (moment 2) of the QSO and neighboring clumps (C1, C2, and C3). The synthesized beam is shown at the bottom-left corner as a white filled ellipse. North is up and east is to the left.}
\label{mom12_qso}
\end{SCfigure*}

\begin{SCfigure*}
\centering
\includegraphics[scale=0.34]{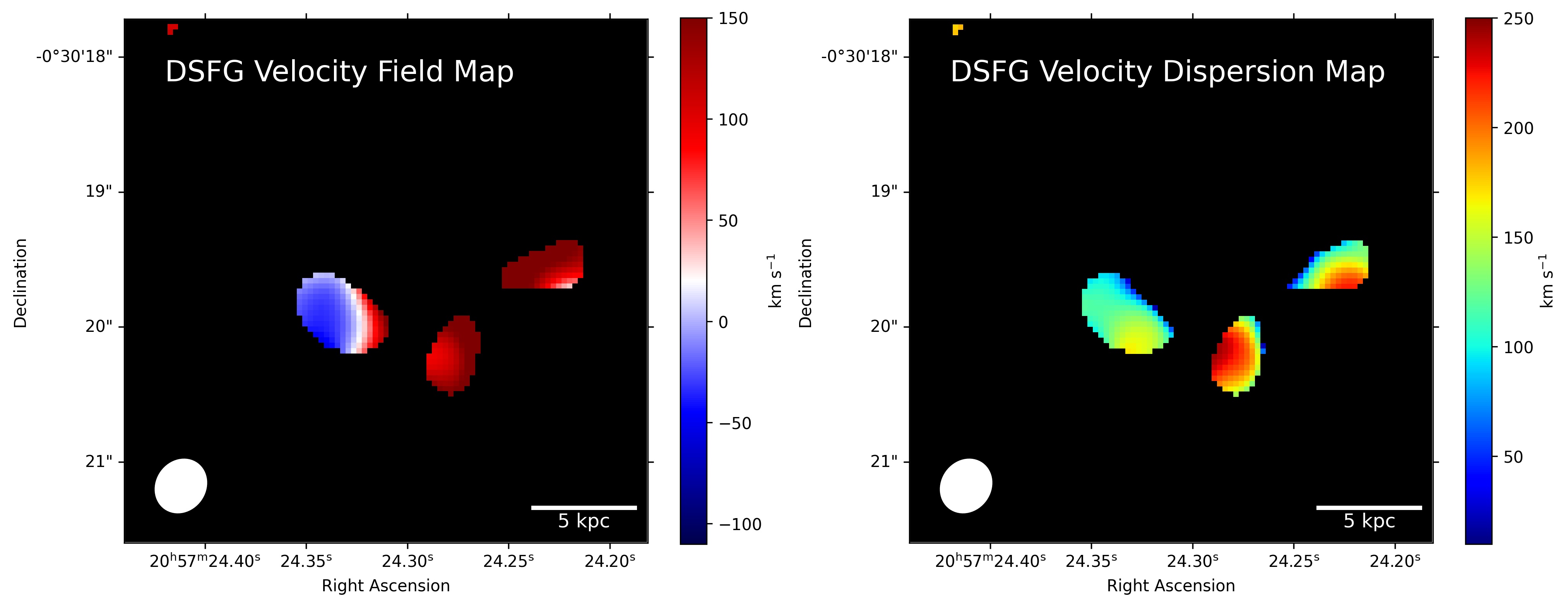}
\caption{Map of the velocity field (moment 1) and velocity dispersion (moment 2) of the DSFG and neighboring clumps (S0 and S1). The synthesized beam is shown at the bottom-left corner as a white filled ellipse. North is up and east is to the left.}
\label{mom12_DSFG}
\end{SCfigure*}

To calculate moment 1 and 2 maps of the QSO and the neighboring clumps, we use the low resolution [C$\textsc{ii}$] line cubes at a rest frame frequency equal to the center of the 1D Gaussian fit applied to the QSO spectra, this is $\nu_0,_{\mathrm[{C\textsc{ii}}]}$ = 334.45 GHz. We do the same with the DSFG and clumps S0 and S1, setting the rest frame frequency equal to the center of the 1D Gaussian fit of the DSFG spectra, $\nu_0,_{\mathrm[C\textsc{ii}]}$ = 334.53 GHz. We do not apply any mask to create these moments but for displaying purposes we show them in Figure \ref{mom12_qso} and \ref{mom12_DSFG} using a mask based on the moment 0, showing the emission at a significance level $\geq$ 2$\sigma$.

Figure \ref{mom12_qso} left panel shows the velocity field map for the QSO and clumps C1, C2, and C3. The QSO presents a marked velocity gradient along the east–west axis, that ranges from $\sim$ -80 km s$^{-1}$ for the eastern region of the host to $\sim$ +50 km s$^{-1}$ in the west side of the QSO, indicative of rotation with an "S" pattern in the center of the galaxy. However this is likely not significant as the region has a size only slightly larger than the synthesized beam. The right panel of Figure \ref{mom12_qso} shows the velocity dispersion map, where the QSO shows a strong dispersion in the northern and southern regions with velocity dispersion of about $\sim$ 180 and $\sim$ 120 km s$^{-1}$, respectively. While these velocities could be due to ongoing physical processes, with the current data we cannot rule out that this pattern may be produced by noise effects due to the low S/N in the borders of the host and beam smearing. For C1, C2, and C3 it is not feasible to derive any conclusion about 
their velocity maps as their sizes are comparable to the beam.

Figure \ref{mom12_DSFG} left shows the velocity field map for the DSFG, S0, and S1. The DSFG exhibits a velocity gradient that ranges from $\sim$ -50 km s$^{-1}$ in the southeast to $\sim$ +80 km s$^{-1}$ in the west, also suggestive of rotation. S0 seems to have a rather homogeneous velocity centered at $\sim$ +100 km s$^{-1}$,
similar to the western part of the DSFG. Found at a projected separation of $\sim$ 6 kpc, this suggests that S0 and the DSFG are physically connected. In the right panel of Figure \ref{mom12_DSFG} we can see that the velocity dispersion is high for S0 and S1. If they are real, they could be interpreted as objects supported by dispersion. However, since both clumps are barely resolved this claim is only speculative. Deeper observations are required to probe their velocity maps.

\begin{SCfigure*}
\centering
\includegraphics[scale=0.34] 
{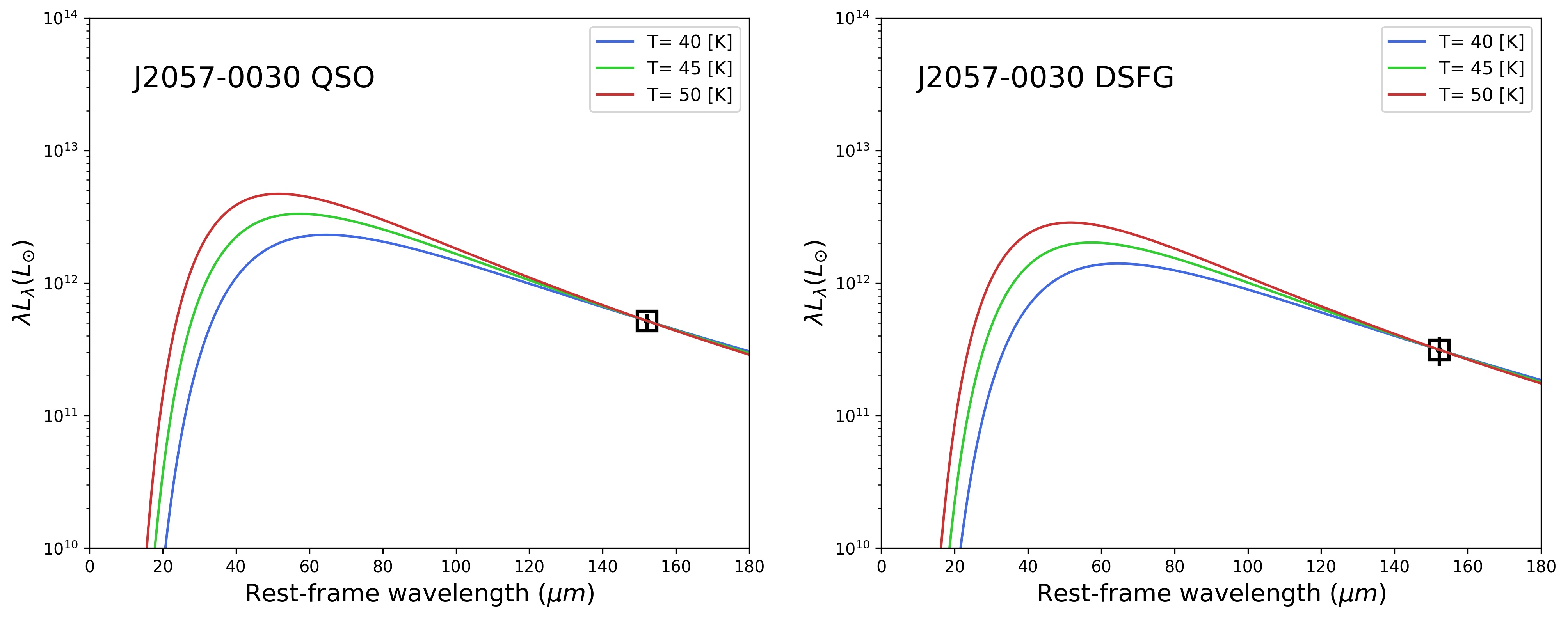}
\caption{FIR SEDs for the QSO and the DSFG using an emissivity coefficient $\beta$ = 1.6 with temperatures of 40, 45, and 50 K represented by the blue, green, and red lines, respectively. The data point corresponds to ALMA detection at 895 $\mu$m (in the observed frame). }
\label{SED}
\end{SCfigure*}

\subsection{SED and SFRs}\label{cap:SEDSFR}

Using the rest frame FIR continuum emission we are able to estimate the total IR emission of our objects. Since the J2057-0030 system only has ALMA detections at $\sim$ 152 $\mu$m, it is not possible to model the spectral energy distribution (SED). Instead, we can provide a well-motivated range that can account for our observations. We assume that in all of our objects the FIR emission is dominated by dust heated through star formation and disregard the AGN contribution to the SEDs as it should be as low as $\sim$ 10$\%$, based on several studies of the mid-to-far IR SEDs of luminous AGN \citep[e.g.,][]{Schweitzer_2006, MorNetzer_2012, Rosario_2012, Lutz_2016}. According to ALMA Cycle 6 Technical Handbook, we consider an additional uncertainty on ALMA absolute flux calibration in band-7 of 10$\%$ and we add this in quadrature to the flux errors. 

For the model, we choose a modified black-body SED to fit the QSO host and the DSFG IR emission using three different temperatures based on two works. First, from  \citet{sommovigo_2022} the range of dust temperatures at $z$ $\sim$ 5 is 35 - 60 K for objects with luminosities on the order of 10$^{11}$ $L_{\odot}$ and SFRs of a few tens of stellar masses per year. This is about one order of magnitude lower than the luminosities and SFRs of the sources in N20, including J2057-0030. Therefore, we consider the objects in \citet{sommovigo_2022} to be representative of a less luminous group of galaxies and we assume that our objects should exhibit temperatures similar or greater than these sources. Thus, we select $T$ = 40 K as a lower limit for the temperature. 

To select an upper limit, we analyze the 9 FIR-bright sources of N20 (detected by Herschel/SPIRE) that represent a more luminous population of objects ($L_{\mathrm{8-1000}}$ $\sim$ 10$^{13}$ $L_{\odot}$) than the J2057-0030 system. For these 9 objects, N20 determined the best gray-body SEDs that fit the 3 Herschel/SPIRE measurements at 250, 350, and 500 $\mu$m and the ALMA observations using a wide range of temperatures $T$ = [40, 45, 50, 55, 60, and 70] K and emissivity coefficients $\beta$ = 1.5 and 1.7. They conclude that 7/9 objects are well fitted by temperatures from 40 K to 50 K (see Figure 7 of N20). Based on this result, we choose a $T$ = 50 K as the upper limit for the temperature. We set an intermediate temperature as the median value of the first two, that is $T$ = 45 K and a $\beta$ = 1.6 \citep[e.g.,][N20]{Beelen_2006, Trakhtenbrot_2017}. 

Figure \ref{SED} shows the SEDs of the QSO and the DSFG, where the blue line represents the SED at $T$ = 40 K, the green at $T$ = 45 K, and the red at $T$ = 50 K. These are the only sources detected in continuum. Meanwhile, the clumps are treated only as upper limits. 
Typically, DSFGs have lower dust temperatures compared to quasars, however, we choose to present our results for the SEDs with the same temperature range for both components even though we expect the DSFG to be biased to the lower bounds, that is $T$ $\sim$ 40 K. 

We estimate the total FIR luminosity ($L_{\mathrm{8-1000}}$) by integrating the resulting SEDs over the 8-1000 $\mu$m range. The $L_{\mathrm{8-1000}}$ of the QSO and the DSFG at $T$ = 45 K is in good agreement with those reported in N20 for SEDs with $T$ = 47 K and $\beta$ = 1.6. 

The SFRs are obtained from these FIR luminosities using the relation from \citet{Kennicutt_1998}, modified as \cite{Nordon_2012} in Equation \ref{SFR}, assuming a Chabrier initial mass function (IMF) from \citet{Chabrier_2003}:

\begin{equation}
\frac{\mathrm{SFR}}{\mathrm{M_{\odot} \ yr^{-1}}} = \frac{1.1 \times L_{\mathrm{IR}}}{10^{10} L_{\odot}}.
\label{SFR}
\end{equation}

The SFRs ranges between a few dozen stellar masses per year in the case of the upper limits for the clumps to a few hundred stellar masses per year in the case of the QSO host and the DSFG (Table \ref{L_SFRs}).

\begin{table*}[htb!]
\centering
\begin{threeparttable}
\caption{Galaxy properties: FIR luminosity and SFR.}
\begin{tabular}{c c c c c c c c c c c}
    \hline
	\hline
	\centering

Component & log $L_{40K}$ & log $L_{45K}$ & log $L_{50K}$ & SFR$_{40K}$ & SFR$_{45K}$ & SFR$_{50K}$ \\
& [$L_{\odot}$] & [$L_{\odot}$] & [$L_{\odot}$]  & [$M_{\odot}$ yr$^{-1}$] & [$M_{\odot}$ yr$^{-1}$] & [$M_{\odot}$ yr$^{-1}$] \\ \hline

QSO & 12.4 $\pm$ 0.06 & 12.56 $\pm$ 0.06 & 12.71 $\pm$ 0.06 & 279 $\pm$ 35 & 402 $\pm$ 51 & 568 $\pm$ 72\\
DSFG & 12.19 $\pm$ 0.11 & 12.35 $\pm$ 0.11 & 12.5 $\pm$ 0.11 & 169 $\pm$ 38 & 244 $\pm$ 38 & 345 $\pm$ 38 \\
C1$^{*}$ & 11.48 $\pm$ 0.33 & 11.64 $\pm$ 0.33 & 11.79 $\pm$ 0.33 & 33 $\pm$ 23 & 48 $\pm$ 23 & 68 $\pm$ 23 \\
C2$^{*}$ & 11.38 $\pm$ 0.34 & 11.54 $\pm$ 0.34 & 11.69 $\pm$ 0.34 & 27 $\pm$ 19 & 38 $\pm$ 19 & 54 $\pm$ 19 \\
C3$^{*}$ & 11.71 $\pm$ 0.19 & 11.87 $\pm$ 0.19 & 12.02 $\pm$ 0.19 & 56 $\pm$ 22 & 81 $\pm$ 22 & 114 $\pm$ 22 \\
C4$^{*}$ & 11.59 $\pm$ 0.3 & 11.75 $\pm$ 0.3 & 11.9 $\pm$ 0.3 & 43 $\pm$ 27 & 62 $\pm$ 27 & 88 $\pm$ 27\\ \hline

\end{tabular}
\vspace{0.25cm} 
\begin{tablenotes} \footnotesize
\item \textbf{Note.} Columns: (1) Component; (2), (3), (4) FIR luminosities (in $L_{\odot}$) calculated by integrating from 8 to 1000 $\mu$m the SEDs constructed with $\beta$ = 1.6 and $T$ = 40 K, 45 K, and 50 K, respectively; (5), (6), (7) star formation rates (in $M_{\odot}$ yr$^{-1}$) estimated using Equation \ref{SFR} with $L_{40K}$, $L_{45K}$ and $L_{50K}$, respectively. Asterisks ($^*$) represent upper limits only.
\end{tablenotes}
\label{L_SFRs}
\end{threeparttable}
\end{table*}

\subsection{$L_{\mathrm{C\normalfont{\textsc{ii}}}}$/$L_{\mathrm{8-1000}}$ versus $L_{\mathrm{8-1000}}$ }\label{cap:Lratio}

 The $L_{\mathrm{C\textsc{ii}}}$/$L_{\mathrm{8-1000}}$ ratio is a measure of the fractional amount of cooling due to gas and dust in star-forming (SF) galaxies. Main sequence (MS) galaxies tend to follow a constant relation, with [C$\textsc{ii}$] luminosities scaling linearly with IR luminosities, although with significant scatter. However, when high-IR luminosity sources such as local (U)LIRGs ($L_{\mathrm{IR}}$ $\geq$ 10$^{12}$ $L_{\odot}$) are analyzed, a decrease in the $L_{\mathrm{C\textsc{ii}}}$/$L_{\mathrm{8-1000}}$ ratio is found, with [C$\textsc{ii}$] being deficient with respect to the IR luminosity. This is also observed in high-$z$ sources \citep[e.g.,][]{Malhotra_1997, DiazSantos_2013, Magdis_2014, DiazSantos_2017}. 

\begin{figure}%[!h]
\centering
\includegraphics[scale=0.38]{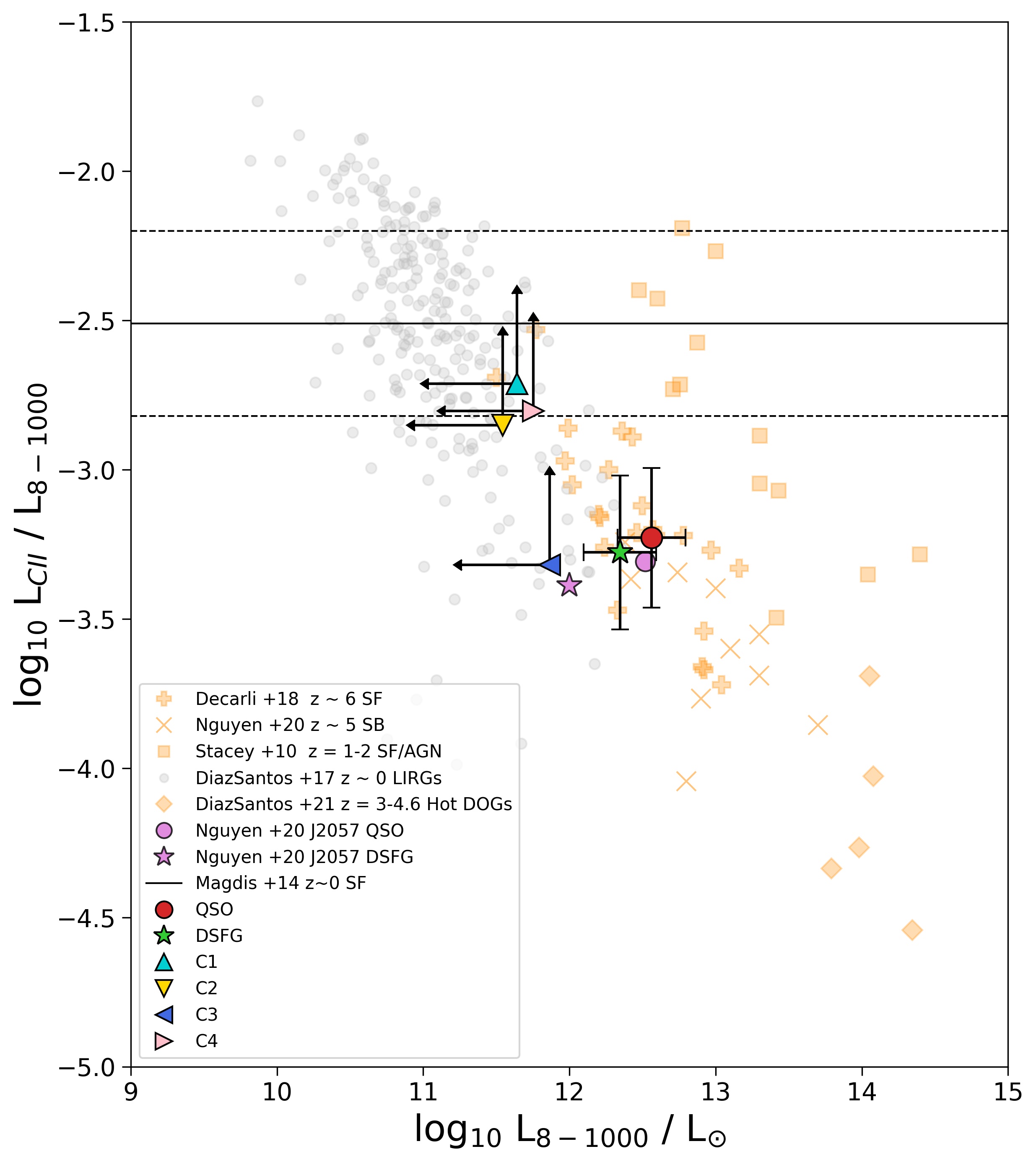}
\caption{$L_{\mathrm{C\textsc{ii}}}$/$L_{\mathrm{8-1000}}$ ratio vs. $L_{\mathrm{8-1000}}$. The sources are plotted as a red circle for the QSO and a green star for the DSFG. The clumps C1, C2, C3, and C4 are plotted as a cyan triangle-up, a yellow triangle-down, a pink triangle-right, and a blue triangle-left, respectively. The $L_{\mathrm{8-1000}}$ are obtained by integrating from 8-1000 $\mu$m the SEDs constructed with $T$ = 45 K and $\beta$ = 1.6. For the clumps, only upper limits of the $L_{\mathrm{8-1000}}$ and lower limits for the $L_{\mathrm{C\textsc{ii}}}$/$L_{\mathrm{8-1000}}$ ratio are shown. The bars represent the quadratic sum of the uncertainties associated with the aperture measurement of the [C$\textsc{ii}$] and FIR emission and the values for $L_{\mathrm{8-1000}}$ and $L_{\mathrm{C\textsc{ii}}}$/$L_{\mathrm{8-1000}}$ obtained using the $L_{\mathrm{8-1000}}$ when the SEDs are constructed with $T$ = 40 K  
and $T$ = 50 K.} 
\label{L_L_e_2T_e2}
\end{figure}

 Figure \ref{L_L_e_2T_e2} shows the [C$\textsc{ii}$] to FIR luminosity ratio as a function of $L_{\mathrm{8-1000}}$. The solid black line represents the relation as estimated by \citet{Magdis_2014} for local normal galaxies from \citet{Malhotra_2001}, which implies a constant $L_{\mathrm{C\textsc{ii}}}$/$L_{\mathrm{8-1000}}$ ratio. The dashed lines represent the scatter in the relation of 0.3 dex. The different components of J2057-0030 are plotted: the QSO (red circle), the DSFG (green star), and the clumps C1, C2, C3, and C4 (as a cyan triangle-up, a yellow triangle-down, a pink triangle-right, and a blue triangle-left, respectively). These points are calculated using the $L_{\mathrm{8-1000}}$ obtained after integrating the SEDs constructed with $T$ = 45 K and $\beta$ = 1.6. For the clumps, only upper limits of the $L_{\mathrm{8-1000}}$ are shown. For comparison, local LIRGS from \citet{DiazSantos_2017} are plotted as gray dots.  High-$z$ sources are plotted as orange symbols: the plus symbols represent SF galaxies from \citet{Decarli_2018}, the squares are galaxies classified as SF and AGN from \citet{Stacey_2010}, the diamonds are hot, dust-obscured galaxies (Hot DOGs) from \citet{DiazSantos_2021}, and the crosses represent the sample from N20. The measurements for the J2057-0030 QSO host and the DSFG obtained by N20 are included as a purple circle and a purple star, respectively. 
 The error bars for our measurements are determined from 
 the uncertainties in the measurement of the [C$\textsc{ii}$] and the FIR emission as follows: for $L_{\mathrm{8-1000}}$ we quadratically add the luminosity ranges obtained using the limiting temperatures described before, that is $T$ = 40 K and $T$ = 50 K.
 Typical $\log (L_{\mathrm{C\textsc{ii}}}$/$L_{\mathrm{8-1000}})$ values for local SF galaxies range from $\sim -2.5$ to $-2$. In contrast, local (U)LIRGs can have ratios about one order of magnitude lower, similar to high-$z$ SF galaxies with $\log (L_{\mathrm{C\textsc{ii}}}$/$L_{\mathrm{8-1000}})$ in the range $\sim -3.5$ to $-2.5$. For high-$z$ sources with $L_{\mathrm{8-1000}}$ $>$ $10^{12}$  $L_{\odot}$ the deficit is very prominent with values of $\log (L_{\mathrm{C\textsc{ii}}}$/$L_{\mathrm{8-1000}})$ $< -3.5$ \citep{Farrah_2013, DiazSantos_2013, DiazSantos_2017, Decarli_2018}. 
The lower limits for our clumps C1, C2, C3, and C4 show $\log (L_{\mathrm{C\textsc{ii}}}$/$L_{\mathrm{8-1000}})$ ratios between $-3.3$ and $-2.8$, closer to local LIRGs ratios \citep[e.g.,][]{DiazSantos_2017}. This result is expected as these components show the lowest FIR continuum. In contrast, the QSO host and the DSFG have $\log (L_{\mathrm{C\textsc{ii}}}$/$L_{\mathrm{8-1000}})$ ratios of $-3.2$ and $-3.3$, respectively. These values are comparable to high-$z$ SF galaxies as those from \citet{Decarli_2018}. From Figure \ref{L_L_e_2T_e2}, it is evident that our sources are [C$\textsc{ii}$] deficient with respect to their FIR luminosity, as observed in local (U)LIRGs with respect to local MS galaxies.  

The physical origin of the [C$\textsc{ii}$] deficit remains unsolved. However, different explanations have been proposed, including reduced photoelectric heating
caused by positively charged dust grains and absorption of ionizing and UV photons by dust in HII regions; optically thick [C$\textsc{ii}$]; the influence of the AGN activity \citep{LangerPineda_2015}; the low metallicity in galaxies \citep{Vallini_2015}; or the [C$\textsc{ii}$] saturation at high temperatures \citep[e.g.,][]{Gullberg_2015}. Whichever the answer, it seems that the deficit is restricted to the nuclear region of galaxies, where the starburst is ongoing \citep{DiazSantos_2014, Smith_2017}. It is important to note that this [C$\textsc{ii}$] deficit 
%at high luminosities
is also reported for other submillimeter lines such as $[\mathrm{O_{I}}]$, $[\mathrm{O_{III}}]$, $[\mathrm{N_{II}}]$ and $[\mathrm{N_{III}}]$ 
indicating that it represents a general effect on most FIR fine structure lines, regardless whether they arise from the ionized or neutral phase of the ISM \citep{GraciaCarpio_2011, Farrah_2013, DiazSantos_2017}.

%--------------------------------------------------------------------

\subsection{Dynamical masses}\label{cap:Mdyn}

We estimate the dynamical masses ($M_{\mathrm{dyn}}$) for the QSO host and the DSFG using the [C$\textsc{ii}$] line, assuming that the [C$\textsc{ii}$]-traced interstellar medium (ISM) is distributed in an inclined, rotating disk \citep[see][]{Wang_2013, Willott_2015} following: 
\begin{equation}
M_{\mathrm{dyn}} = 9.8\times10^8 \left(\frac{D_{[\mathrm{C\textsc{ii}}]}}{\mathrm{kpc}}\right) \left[\frac{{\mathrm{FWHM}[\mathrm{C\textsc{ii}}]}}{100 \ \mathrm{km s^{-1}}}\right]^2 \frac{1}{\sin^2(i)} M_{\odot},
\label{Mdyn}
\end{equation}

\noindent where $D_{[\mathrm{C\textsc{ii}}]}$ is the size of the [C$\textsc{ii}$] emitting
region measured by the deconvolved major axis of the 2D
Gaussian fit. The term $\sin(i)$
represents the inclination angle between the line of sight and the polar axis of the host gas disks, with the circular velocity given as $v_{circ}$ = 0.75 $\times$ FWHM/$\sin(i)$. Assuming a thin-disk geometry, we can determine the inclination of the disk ($i$) following cos(i) $= a_{\mathrm{min}}$/$a_{\mathrm{maj}}$, where $a_{\mathrm{min}}$, $a_{\mathrm{maj}}$ are the semi-minor and semi-major axes of the [C$\textsc{ii}$] emitting regions, respectively. The uncertainties for this method are large and driven by the measurements of the major and minor axis of the [C$\textsc{ii}$] emitting region, which are used to estimate D$_{[\mathrm{C\textsc{ii}}]}$ and $i$. The dynamical masses are reported in Table \ref{Masses}, where we have also included the inclination uncorrected masses $M_{\mathrm{dyn}}^{\mathrm{uncorr}}$ (i.e., $M_{\mathrm{dyn}}^{\mathrm{uncorr}}$ = $M_{\mathrm{dyn}}$ $\times$ $\sin^2$(i)). For the QSO host $M_{\mathrm{dyn}}$ is 10$^{11.4 \pm 0.1}$ $M_{\odot}$ and $M_{\mathrm{dyn}}^{\mathrm{uncorr}}$ $\approx$ 10$^{10.9 \pm 0.1}$ $M_{\odot}$. For the DSFG $M_{\mathrm{dyn}}$ and $M_{\mathrm{dyn}}^{\mathrm{uncorr}}$ are $\approx$ 10$^{11 \pm 0.1}$ $M_{\odot}$.

Despite the fact that our sources show clear rotation disks, the random motions in both galaxies is significant, with dispersion velocities of $\sim$ 150 km s$^{-1}$ in the central region of the QSO and $\sim$ 130 km s$^{-1}$ in the DSFG. This motivates the assumption that our galaxies are not supported by rotation but instead they are dominated by dispersion as in the case of the galaxy bulge \citep{Decarli_2018}. In this case, we can estimate the dispersion mass ($M_{\mathrm{disp}}$) using the following expression, derived from the virial theorem: 
\begin{equation}
M_{\mathrm{disp}} = \frac{3}{2} \frac{a_{\mathrm{maj}}\sigma_{\mathrm{line}}^2}{G} ,
\label{Mdisp}
\end{equation}

\noindent where $\sigma_{\mathrm{line}}$ is the line width of the Gaussian fit to the [C$\textsc{ii}$] spectrum and G is the gravitational constant. These dispersion masses are $M_{\mathrm{disp}}$ $\approx$ 10$^{10.5 \pm 0.1}$ $M_{\odot}$ for the QSO host and the DSFG (see Table \ref{Masses}).

The uncertainties in the masses of about 0.1 dex are rather small 
as they include only formal, measurement related errors, without taking into account the uncertainties in the model used to describe the kinematics of the system, which are expected to be large based on how different the dynamical masses are when we assume a disk model ($M_{\mathrm{dyn}}$) compared to a bulge model ($M_{\mathrm{disp}}$). 

\subsection{Dust masses $M_{\mathrm{dust}}$}\label{cap:Mdust}

The continuum emission at rest wavelength $\lambda$ $\sim$ 152 $\mu$m can be used to calculate the dust masses $(M_{\mathrm{dust}})$ of our QSO host and its companion DSFG assuming the FIR continuum originates from optically thin dust at these wavelengths \citep{Dunne_2000, Beelen_2006}, using:
\begin{equation}
M_{\mathrm{dust}} = \frac{S_{\lambda}(\lambda_{\mathrm{rest}}) D_L^2}{\kappa_d(\lambda_{\mathrm{rest}})B_{\lambda}(\lambda_{\mathrm{rest}}, T_d)} ,
\label{Mdust}
\end{equation}

\noindent where $S_{\lambda}(\lambda_{\mathrm{rest}})$ is the continuum flux density at $\lambda_{\mathrm{rest}}$, $D_L$ is the luminosity distance, $\kappa_d(\lambda_{\mathrm{rest}})$ is the wavelength dependent dust mass opacity and $B_{\lambda}(\lambda_{\mathrm{rest}}, T_d)$ is the monochromatic value of the Planck function at $\lambda_{\mathrm{rest}}$ for the dust temperature $T_d$. For $\kappa_d$ at 850 $\mu$m, \citet{Dunne_2000} assumed a value of   0.077 m$^2$ kg$^{-1}$, then $\kappa_d$ can be calculated as $\kappa_d = 0.077\times (850/\lambda_{\mathrm{rest}})^\beta$ m$^2$ kg$^{-1}$. We use $T_d$ = 45 K and $\beta$ = 1.6, the same values used for the SED analysis. We need to be aware that since SFRs and $M_{\mathrm{dust}}$ are estimated from the same measurements of continuum emission, both quantities are strongly correlated.

For the QSO host $M_{\mathrm{dust}}$ $\sim$ 10$^{8.5 \pm 0.4}$ $M_{\odot}$ 
and for the DSFG $M_{\mathrm{dust}}$ $\sim$ 10$^{8.3 \pm 0.4}$ $M_{\odot}$ (see Table \ref{Masses}). 
The uncertainties in the dust masses include the errors in the measurements of the continuum emission and the errors based on the literature for the dust temperature and the dust emissivity coefficient $\beta$. The uncertainties are set to $\Delta{T_d} = 5$ K \citep{sommovigo_2022}
and $\Delta{\beta}$ = 0.5 \citep[e.g.,][]{DaCunha_2021, Algera_2023}. The contribution to the errors from the continuum emission and the $T_d$ uncertainties have a small impact compared to the uncertainties from $\beta$.

\begin{table*}
\centering
\begin{threeparttable}
\caption{Galaxy properties: masses.}
\begin{tabular}{c c c c c c c}
	\hline
	\hline
	\centering

Component & log $M_{\mathrm{dyn}}^{\mathrm{uncorr}}$ & log $M_{\mathrm{dyn}}$ & log $M_{\mathrm{disp}}$ & log $M_{\mathrm{dust}}$ & log $M_{\mathrm{gas}}$ & log $M_{\mathrm{mol}}$ \\
 & [$M_{\odot}$] &  [$M_{\odot}$] & [$M_{\odot}$] & [$M_{\odot}$] & [$M_{\odot}$] & [$M_{\odot}$] \\ \hline

QSO Host & 10.9 $\pm$ 0.1 & 11.4 $\pm$ 0.1 & 10.5 $\pm$ 0.1 & 8.5 $\pm$ 0.4 & 10.5 $\pm$ 0.4 & 10.9 $\pm$ 0.3\\
DSFG & 11.0 $\pm$ 0.1 & 11.0 $\pm$ 0.1 & 10.5 $\pm$ 0.1 & 8.3 $\pm$ 0.4 & 10.3 $\pm$ 0.4 & 10.5 $\pm$ 0.3 \\
 \hline

\end{tabular}
\vspace{0.25cm} 
\begin{tablenotes} \footnotesize
\item \textbf{Note.} Columns: (1) Component; (2) Uncorrected dynamical mass ($M_{\mathrm{dyn}}^{\mathrm{uncorr}}$ = $M_{\mathrm{dyn}}$ $\times$ $\sin^2$(i)); (3) Dynamical mass ($M_{\mathrm{dyn}}$) estimated using Equation \ref{Mdyn}; (4) Dispersion mass ($M_{\mathrm{disp}}$) estimated using Equation \ref{Mdisp}; (5) Dust mass ($M_{\mathrm{dust}}$) estimated using Equation \ref{Mdust}; (6) Gas mass ($M_{\mathrm{gas}}$) derived from $M_{\mathrm{dust}}$ and a GDR; (7) Molecular gas mass ($M_{\mathrm{mol}}$) derived from [C$\textsc{ii}$] using Equation \ref{Zanella} from \citet{Zanella_2018}. All masses are reported in $M_{\odot}$. 
\end{tablenotes}
\label{Masses}
\end{threeparttable}
\end{table*}

\subsection{Gas masses $M_{\mathrm{gas}}$}\label{cap:Mgas}

Molecular gas (H$_2$) is the most abundant molecule in the ISM but is difficult to detect in emission due to its lack of a permanent dipole moment and the high temperatures necessary to excite the rotational transitions, which means that direct observations of H$_2$ can only trace a small and warm (T $>$ 100 K) fraction of the total H$_2$. Commonly, CO has been used as a tracer of cold H$_2$, as well as the FIR/submillimeter dust continuum, and recently, [C$\textsc{ii}$].

In this context, we determine the gas masses of our galaxies following two different methods. First, we calculate the total $M_{\mathrm{gas}}$ (atomic + molecular) using the dust mass obtained from the FIR continuum \citep[see][]{Magdis_2012, Santini_2014, Genzel_2015, Scoville_2017} with a gas-to-dust ratio (GDR) of 100 as used in N20, based on the \citet{Draine_2007} determination of GDR for local, solar metallicity galaxies. The GDR can vary significantly among high-$z$ galaxies, with studies showing a wide range of values \citetext{\citealp{Ivison_2010} with a GDR $\sim$ 40 at $z = 2.3$, \citealp{Banerji_2016} finding a GDR $\sim 30 - 110$ at $z$ $\sim 2.5$}. These lower ratios are unexpected as the GDR scales inversely with metallicity \citep{Leroy_2011, Remy_2014} and high-$z$ sources are known to have lower metallicities than local galaxies. This discrepancy can be a result of the methodology used in those works to determine the ratio as it is based on dust measurements and gas masses estimated from CO, which requires a value for the CO-to-H$_2$ conversion factor ($\alpha_{\mathrm{CO}}$), that is highly dependant on the properties of the galaxy, including its metallicity both in local galaxies \citep{Leroy_2011} and high-$z$ galaxies \citep{Heintz_2021, Sanders_2022e}. Because of this, we choose to use a GDR = 100, but it has to be noticed that our mass estimates will scale linearly with any revised value of this ratio.

The estimated gas mass for our sources is $M_{\mathrm{gas}}$ $\sim$ 10$^{10.5 \pm 0.4}$ $M_{\odot}$ for the QSO host and $\sim$10$^{10.3 \pm 0.4}$ $M_{\odot}$ for the DSFG. The uncertainties in the $M_{\mathrm{gas}}$ are based on the errors in the dust masses estimations only and therefore they are expected to be larger if we include the uncertainties of the potentially different GDR.

The other method is based on the correlation between [C$\textsc{ii}$] and the molecular gas mass ($M_{\mathrm{mol}}$) in galaxies, as discussed in several works 
\citep[e.g.,][]{Zanella_2018, Madden_2020, Dessauges_2020, Vizgan_2022}. Particularly, at high-$z$ the determination of the molecular gas content becomes challenging. As the metallicity of the galaxies decreases with redshift, common tracers such as CO and dust are not as efficient as they are in the local universe since dust abundances decrease and the CO molecule is dissociated and ionized into C and C$^+$, while the H$_2$ molecule is either dust or self-shielded from UV radiation leading to a fraction of H$_2$ that is not being traced by carbon monoxide (also known as CO-dark gas).

In particular, \citet{Zanella_2018} (hereafter Z18) have shown that [C$\textsc{ii}$] is a convenient tracer of H$_2$ regardless of the galaxy nature, from main-sequence (MS) to starburst galaxies, including metal-poor systems. Using a sample covering a redshift range of 0 $<$ $z$ $<$ 6 and metallicities in the range 7.9 $<$ 12 $+$ log(O/H) $<$ 8.8 they find that [C$\textsc{ii}$] luminosities correlate tightly with molecular gas masses in a (nearly) linear relation following:
\begin{equation}
\log L_{\mathrm{C\textsc{ii}}} = -1.28 (\pm 0.21) + 0.98 (\pm 0.02) \log M_{\mathrm{mol}} ,
\label{Zanella}
\end{equation}

\noindent with a dispersion of 0.3 dex. Nonetheless, the Z18 relation is derived from a heterogeneous sample of galaxies found in the literature, and with measurements of [C$\textsc{ii}$] and $M_{\mathrm{mol}}$ estimated from CO or dust continuum. Also, these quantities are not calibrated to account for the fraction of CO-dark molecular gas in lower metallicity galaxies. 

At any rate, using the Z18 relation we estimate $M_{\mathrm{mol}}$ $\sim$ 10$^{10.9 \pm 0.3}$ $M_{\odot}$ for the QSO host and $M_{\mathrm{mol}}$ $\sim$ 10$^{10.5 \pm 0.3}$ $M_{\odot}$ for the DSFG, with uncertainties similar to the dispersion reported by the Z18 fit, showing that our errors are dominated by the relation itself rather than by the uncertainties in $L_{\mathrm{C\textsc{ii}}}$. Both masses are larger than those determined using the dust masses and a GDR $=100$.

\subsection{The main sequence of galaxies}\label{cap:MS}

The relation between the SFR and the stellar mass ($M_{\star}$) of galaxies is known as the star-forming main sequence of the galaxies (MS). 
Observations at $z$ $<$ 4 show that the MS seems to have a nearly constant slope, close to unity \citep{Tomczak_2016, Santini_2017}. However, the MS needs to be normalized to account for the increase of star formation per unit stellar mass as the redshift increases. This normalization depends on the sample selection and SFR tracer used \citep{Speagle_2014}. 

In this work, we study the position of QSO host and the DSFG relative to the MS relation for high-$z$ sources. We use the gas masses derived from dust ($M_{\mathrm{gas}}$) and the molecular gas mass ($M_{\mathrm{mol}}$) using Equation \ref{Zanella} (following Z18) to determine their stellar masses. 

Assuming that the total dynamical mass is $M_{\mathrm{dyn}} = M_{\mathrm{gas}} + M_{\star}$ (i.e., the dust mass is negligible and dark matter is not dominant within the galactic radius where the dust and gas emission is observed) we can estimate $M_{\star}$ by subtracting the gas mass from the total (inclination corrected) dynamical mass of each source. We see that both of our sources have $M_{\mathrm{gas}} < M_{\mathrm{dyn}}$, indicating that the adopted GDR is plausible for our sources. The gas masses estimated using $M_{\mathrm{dust}}$ are consistent with a gas content $M_{\mathrm{gas}} \approx 0.2 \times M_{\star}$ for the QSO host and $M_{\mathrm{gas}} \approx 0.25 \times M_{\star}$ for the DSFG. The gas content using $M_{\mathrm{mol}}$ from $L_{\mathrm{C\textsc{ii}}}$ is $M_{\mathrm{gas}} = 0.4 \times M_{\star}$ for the QSO host and $M_{\mathrm{gas}} = 0.5 \times M_{\star}$ for the DSFG. These values are in good agreement with the gas fractions found in the literature, corresponding to $\sim 0.5 - 0.8$ and determined for non-active galaxies out to $z \sim$ 4 \citep{Schinnerer_2016, Dessauges_2017, Darvish_2018, Gowardhan_2019}.

\begin{figure}%[!h]
\centering
\includegraphics[scale=0.37]{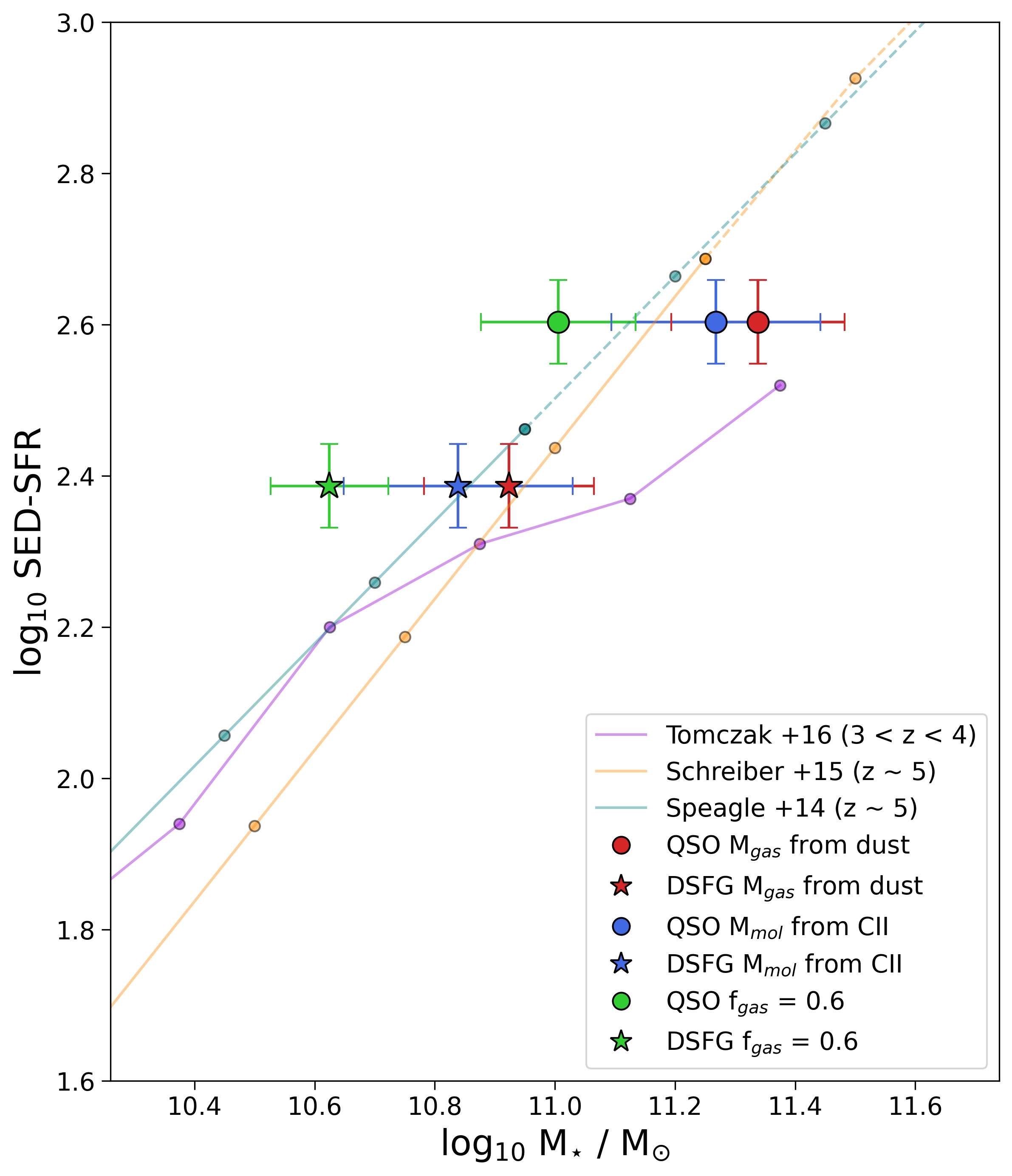}
\caption{Stellar mass vs. star formation rate (main sequence of star-forming galaxies). The filled circle and star represent the QSO host and the DSFG, respectively. We estimate the stellar mass using $M_{\star}$ = (1 - $f_{\mathrm{gas}}$) $M_{\mathrm{dyn}}$ with $f_{\mathrm{gas}}$ $\equiv$ $M_{\mathrm{gas}}$/$M_{\mathrm{dyn}}$. When $M_{\mathrm{gas}}$ is calculated using the dust mass $M_{\mathrm{dust}}$ and a gas-to-dust ratio (GDR = 100) the markers are colored red. When the $M_{\mathrm{mol}}$ comes from \citet{Zanella_2018} the markers are blue. For the case $M_{\star}$ = 0.4 $M_{\mathrm{dyn}}$ \citep[as done in][]{Nguyen_2020} the markers are green. For comparison, three MS curves for high-$z$ sources are plotted: \citet{Speagle_2014}, \citet{Schreiber_2015} and \citet{Tomczak_2016}.}
\label{MS}
\end{figure}

Figure \ref{MS} shows the position of our sources on the MS plane adopting stellar mass derived from dust and [C$\textsc{ii}$], as described above. We also plot the stellar mass of the QSO host and the DSFG as estimated in N20 assuming $f_{\mathrm{gas}}$ = 0.6. We call this our third stellar mass determination. We include MS relations taken from \citet{Speagle_2014} based on a compilation of 25 studies of galaxies at $0.16 < z < 3.15$, after correcting the IMF from Kroupa to Chabrier for consistency. The second MS relation is the parametrization from Equation 9 of \citet{Schreiber_2015} for galaxies at redshifts $4 < z < 5$, after correcting the SFR and M$_{\star}$ from Salpeter to Chabrier IMF. The last one is the \citet{Tomczak_2016} MS relation for galaxies at $3 < z < 4$.   

The QSO host seems to be
laying close to the MS curves from \citet{Speagle_2014} and \citet{Schreiber_2015} if we expand these relations to higher masses than those used in their works (Figure \ref{MS}, dotted lines). If we compare with \citet{Tomczak_2016}, it seems to be above the right end of the curve ($M_{\star}$ $\sim$ 10$^{11}$ $M_{\odot}$), which is in agreement with N20, where FIR-Faint sources seem to be above the MS relations of \citet{Schreiber_2015} and \citet{Tomczak_2016}. For the DSFG the three methods are in good agreement and the position of the galaxy is on or close to all the MS relations plotted, suggesting that the DSFG is likely a MS galaxy at this cosmic time. This result is robust considering the formal uncertainties for $M_{\star}$, but as previously discussed, they are expected to be even larger. 

%--------------------------------------------------------------------

\section{Discussion} \label{Chapter5}

Based on ALMA Cycle 6 observations, the J2057-0030 system can be described as composed of a QSO and a DSFG, surrounded by several smaller structures, which are undetected in continuum emission. These clumps support the idea that we are witnessing a strongly interacting, where the sources have been perturbed by gravitational forces. This is supported by our measurements, with the QSO and the DSFG having a projected spatial separation of $\sim$ 20 kpc 
with a small velocity line of sight offset of $\Delta{v}$ = 68 km s$^{-1}$. This result is in agreement with merger systems observed at high-$z$ \citep[e.g.,][]{Banados_2018, Ginolfi_2020, Romano_2021} . 

Focusing on the tail formed by C1, C2, and C3 with a length of $\sim$ 10 kpc, we consider it unlikely to represent a cold stream accretion from the CGM or IGM. As the tail is detected in [C$\textsc{ii}$], it suggests that this structure is already metal enriched and does not represent pristine gas being accreted from the cosmic web. Still, it is possible to observe diffuse metal-rich gas in the circumgalactic medium, up to diameter-scales of 10-30 kpc \citep[see][]{Fujimoto_2020, Ginolfi_2020, DiCesare_2024}. This metal enrichment may be caused by past episodes of outflows \citep{Pizzati_2020} and gas stripping \citep{Ginolfi_2020}. Moreover, the intergalactic transfer of metal-rich gas via streams, in massive systems as J2057-0030, may represent a dominant channel for mass assembly, as seen in simulations \citep[e.g.,][]{AnglesAlcazar_2017} and observations \citep[e.g.,][]{DiazSantos_2018}. On the other hand, some theoretical expectations of cold flows predict large amounts of angular momentum, which is inconsistent with the narrow velocity range of the tail. This is also in disagreement with the outflows reported in the literature where broad [C$\textsc{ii}$] wings are seen \citep[e.g.,][]{Stanley_2019, Bischetti_2019}. In addition, the different kinematics observed in the clumps, as suggested by their different line profiles, indicate that they are indeed individual structures and they do not represent a continuous stream of gas being accreted.

Instead, we suggest that the tail is a tidal stream from a major merger between the QSO and the DSFG. Tidal tails are commonly observed in local mergers, where usually one tail emanates from each galaxy involved (e.g., NGC 4676 "The Mice"; Arp 295; M51 $+$ NGC 5195; NGC 4038/9 "The Antennae"). However, there are also observations of single tails similar to the one observed in J2057-0030 at low-$z$ (e.g., Arp 100, 101, 173, 174). The single tail in J2057-0030 is particularly bright in [C$\textsc{ii}$] with an extension of 10 kpc. Such a prominent tidal feature may represent a prograde interaction, where the galactic spin is aligned with the angular momentum of the orbit as seen in simulations \citep{Privon_2013}. The absence of a "counterpart tail" could be explained by the DSFG having a retrograde spin or a dynamically "warmer" disk (i.e., less ordered rotation pattern). 

The undetected continuum of the clumps can be explained as similar to tidal tails observed at low-$z$, that usually correspond to material pulled out from the outer regions of the merging galaxies, where the metallicity tends to be lower and therefore resulting in lower FIR emission. Regarding the geometry of the tail, it could be a projection effect of a curved tail seen almost edge-on, in which case there should be a velocity gradient along the tail that we cannot confirm. In fact, the spectral signatures of the three clumps (C1, C2, C3) do not appear to show a systematic velocity shift, but instead complex kinematics with embedded substructures. 

In this merger scenario, we could also expect a tidal bridge connecting the QSO and the DSFG. We propose that S0 and S1 are indeed "real" sources that represent the bridge between our QSO host and its companion. If this is not the case, we could explain this missing feature as being fainter enough to be below our detection threshold or so short-lived that it has already been disrupted or accreted. Hence, even if the bridge is not present, the hypothesis of a merger is still highly likely. Our system could represent a high-$z$ analog to the systems Arp 101 or Arp 173.

We draw attention to C4, located at $\sim$ 28 kpc from the QSO and $\sim$ 42 kpc from the DSFG. The presence of this clump, at the same redshift of the two main galaxies, shows that the sources associated with J2057-0030 span at least 50 kpc in size. This is comparable to the most extended systems known to date, such as WISE J224607.56-052634.9 \citep{DiazSantos_2018}, a multiple-merger system at $z \sim 4.6$ that hosts the most luminous IR quasar known, Candels-5001, \citep{Ginolfi_2017}, a starburst galaxy at $z \sim 3.4$ with a $\sim 40$ kpc massive molecular gas cloud, or cid$\_$346 \citep{Cicone_2021}, an apparently isolated unobscured AGN at $z \sim 2.2$ surrounded by a $\sim 200$ kpc CO reservoir. Most $z > 2$ quasars, however, report modest, $\lesssim 10$ kpc \citep{Klamer_04,Fujimoto_2019,Fujimoto_2020,Herrera_Camus2021,Akins_2022,Fogasy_2022} or very compact sizes \citep{Polleta_2011,Riechers_2011,Decarli_2018,Novak_2020,Fogasy_2020} for their circumnuclear medium. Besides, the SFRs found in J2057-0030 are below the most extreme examples ($40 - 400 M_{\odot}$ yr$^{-1}$), rendering it on or close to the main sequence of star-forming galaxies. 

In the model proposed by \cite{Sanders_1988}, obscured quasars are mainly fueled by major mergers between gas-rich galaxies that cause episodes of rapid starburst and SMBH growth. During this phase, most of the accretion onto the central region of galaxies is expected to occur while the SMBH is enshrouded by gas and dust, making it severely obscured \citep[e.g.,][]{Hopkins_2006}. After this period, the circumnuclear gas and dust are expected to be expelled from the galaxy and the SMBH surroundings by the stellar or quasar feedback, causing a significant decrease in the SFR. This leads to an unobscured phase for the quasar after the merging ends \citep{Treister_2010}. However, it is still debated whether the Sanders picture holds as a evolutionary sequence that explains quasar activity: while a recent compilation shows no correlation between mergers and quasar activity \citep{Villforth_2023}, the largest study so far finds that such correlation exists \citep{Goulding_2017}. In fact, the merger state of our source, contradicts this simplistic scenario by being one of the brightest unobscured quasars at $z \sim 5$  with evidence of important star formation while the merger event is still ongoing. This is indicated by the presence of a prominent tidal tail, which is expected to be formed after the first pericenter or first close passage between the interacting galaxies, as seen in simulations \citep[e.g.,][]{Privon_2013, Volonteri_2015, Prieto_2021}. In other words, J2057-0030 seems to combine all the stages described by the evolutionary model of \cite{Sanders_1988} in a single cosmological snap-shot. 

The emerging picture is that high-$z$ quasars reside in high-density environments. The evidence comes from the fraction of high-$z$ quasars with  close companions and by the clustering of galaxies in larger field searches around quasars \citep{Venemans_2020, GarciaVergara_2019}. Whether or not mergers play a role in the AGN activity, still remains unclear. However, it is now becoming evident that most of the non-active galaxies in these environments are SMGs and DSFGs with very high levels of obscuration \citep[e.g.,][]{Thomas_2023}. This could explain the very different level of clustering around quasars of optically and  identified galaxies \citep{GarciaVergara_2022}. It is intriguing that even though  selected SMG and DSFG samples show varying levels of obscuration, in many cases allowing for an optical and/or NIR counterpart to be observed \citep{Chen_2015, Chen_2016} and even allowing for optical spectroscopic follow up \citep{Danielson_2017}, quasars companions usually correspond to completely enshrouded sources. Whether the enhanced obscuration in these environments is due to an evolutionary phase or to the still to be fully characterised nature of systems at the highest densities at $z \sim$ 5-6, or both, it is something to be determined in the future.
%--------------------------------------------------------------------

\section{Conclusions}\label{Chapter6}

We have presented ALMA Cycle 6 data and analysis for the unique J2057-0030 system composed of a QSO at $z \sim 4.68$, a companion DSFG, and including a clumpy circumgalactic medium, all of which extend for at least 50 kpc in size. The observations include the [C$\textsc{ii}$] line 
and underlying FIR continuum emission.

\begin{enumerate}

\item The system is composed of the QSO, the DSFG and the extended [C$\textsc{ii}$] emission described as a "tail", 
formed by three different "clumps": C1, C2, and C3. Two other sources are discovered close to the DSFG (S0 and S1). A final distant clump C4 is observed in the SW direction from the QSO. The QSO and the DSFG are detected in [C$\textsc{ii}$] and continuum meanwhile all the clumps remain undetected in FIR continuum. 

\item The [C$\textsc{ii}$] 1D spectra reveals that the QSO host and the DSFG have a velocity offset of only $\Delta{v}$ = 68 km s$^{-1}$, which combined with the projected spatial separation of $\sim$ 20 kpc represent a strong indication that both sources are gravitationally bound. These values are comparable to other interacting galaxies observed at similar redshift. This result and the presence of several [C$\textsc{ii}$] clumps lead us to propose that the J2057-0030 system is probably undergoing a merger event. 

\item The FIR luminosities are $L_{\mathrm{8-1000}}$ $\sim$ $10^{12.6}$ $L_{\odot}$ for the QSO host and $L_{\mathrm{8-1000}}$ $\sim$ $10^{12.4}$ $L_{\odot}$ for the DSFG.
yielding SFRs of 400 $M_{\odot}$ yr$^{-1}$ for the QSO host and 240 $M_{\odot}$ yr$^{-1}$ for the DSFG. We estimate the [C$\textsc{ii}$] to FIR luminosity ratio for the two main sources while reporting the lower limits for the clumps (Figure \ref{L_L_e_2T_e2}). The [C$\textsc{ii}$] deficits observed in the QSO and the DSFG are comparable to deficits seen in normal SF galaxies at high-$z$. The clumps are consistent with values determined from local (U)LIRGs reported in the literature.

\item We estimate rough values for the dynamical, dust, and gas masses for our main sources. 
The stellar masses are estimated to be $\approx$ 10$^{11.2}$ and 10$^{10}$ $M_{\odot}$ which locate the QSO host and the DSFG on the main sequence of star-forming galaxies when compared to high-$z$ MS curves (Figure \ref{MS}). This result is intriguing considering that the J2057-0030 unobscured QSO was selected from a sample containing the brightest quasars at this redshifit, however its host galaxy shows a modest SFR. 

\item The extended morphological properties observed in J2057-0030, suggest that the  merger between the QSO and the DSFG is somewhere after the first pericenter and the latest stage of merging, with a modest SFRs observed in both galaxies. This contradicts the current model of quasar evolution where mergers play a significant role in the growth of their BHs and host galaxies. However, this is likely not an uncommon scenario but rather another example of a high-$z$ unosbucured QSO in a dense environment, where the companion is a non-active obscured galaxy on its way to merge. Whether the enhanced obscuration in quasar environments is due to an evolutionary phase or to the nature of systems at high-$z$ is not yet clear.

\end{enumerate}

Based on our results, we consider that deeper observations of the J2057-0030 system using ALMA and JWST would allow us to resolve the QSO host and DSFG and have a proper measurement of their properties (e.g., stellar mass, kinematics, circumnuclear region). We could be able to model the disks of our galaxies and compare their morphology and physical properties with simulations of mergers at high-$z$. This could allow us to explore the complexity of the interplay between galaxies and AGN when the Universe was only a fraction of its current age.\\

{\bf Acknowledgements:} The authors acknowledge support from the National Agency for Research and Development (ANID) FONDECYT Regular 1201748 (MFF, PL). TDS was supported by the Hellenic Foundation for Research and Innovation (HFRI) under the "2nd Call for HFRI Research Projects to support Faculty Members $\&$ Researchers" (Project Number: 3382). BT acknowledges support from the European Research Council (ERC) under the European Union’s Horizon 2020 research and innovation program (grant agreement number 950533) and from the Israel Science Foundation (grant number 1849/19). This paper uses the following ALMA data: ADS/JAO.ALMA$\#$2018.1.01830.S. ALMA is a partnership of ESO (representing its member states), NSF (USA) and NINS (Japan), together with NRC (Canada), MOST and ASIAA (Taiwan), and KASI (Republic of Korea), in cooperation with the Republic of Chile. The Joint ALMA Observatory is operated by ESO, AUI/NRAO and NAOJ.

\bibliographystyle{aa}
\bibliography{bibs}

\appendix
\section{Calibration Issues}\label{cap:Calibrations}

\begin{figure}[!h]
\centering
\includegraphics[scale=0.29]{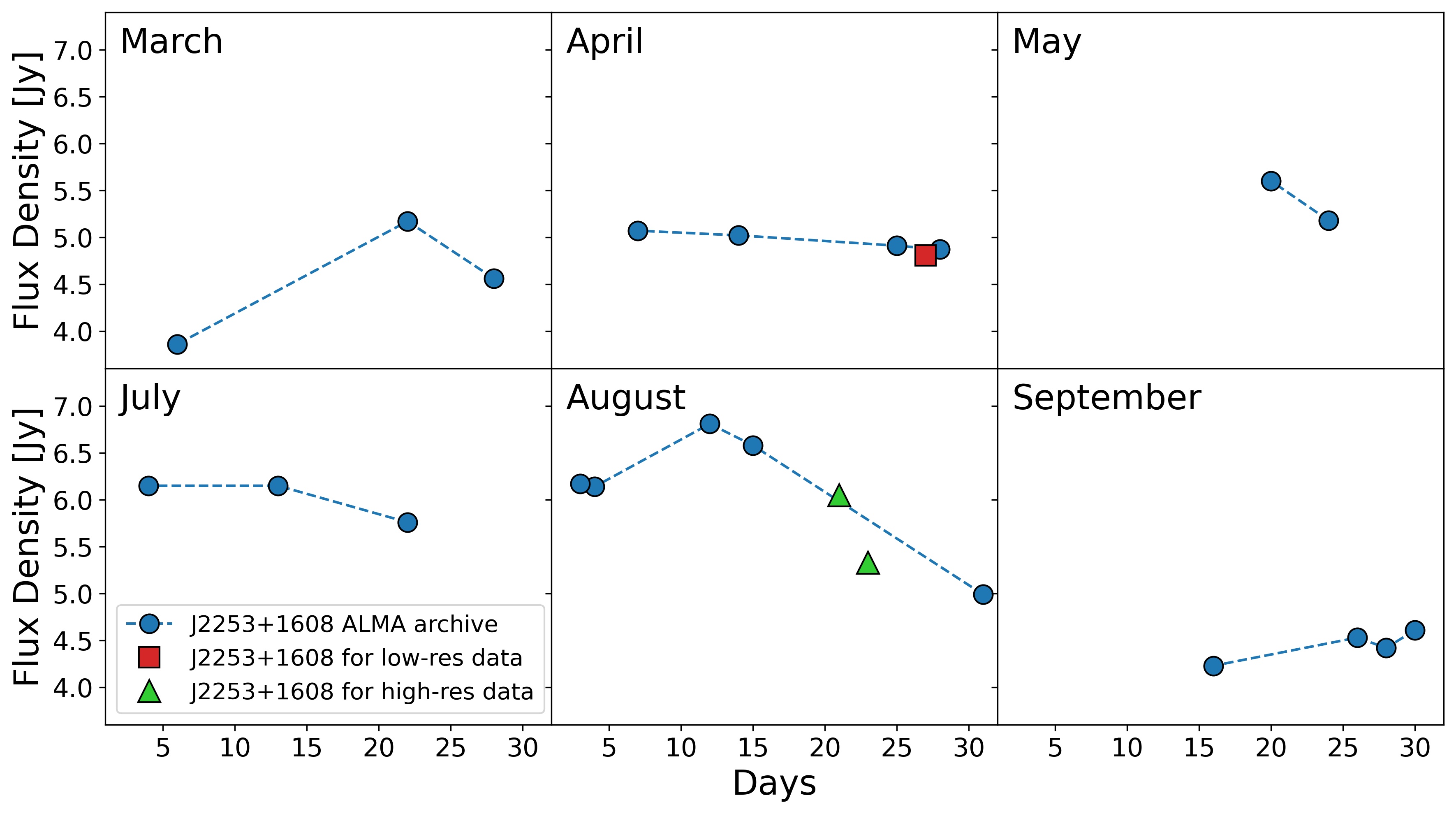}
\caption{J2253+1608 fluxes from March to May and July to September in 2019 as reported by ALMA. The red square symbolizes the J2253+1608 flux used to calibrate the low resolution data, taken on April 27. The green triangles symbolize the J2253+1608 flux used to calibrate each MS of the high resolution data, taken on August 21 and 23.}
\label{calibrator}
\end{figure}

Visual inspection of the [C$\textsc{ii}$] spectra showed significant differences in peak and integrated fluxes when comparing the combined high and low resolution data with the individual data sets at different resolutions, with higher fluxes found in the combined data cube. The same issue was observed in the continuum fluxes. We explored if this was a result of the incorrect flux scaling described in \citet{JvM_1995} (the "JvM" effect), which is caused by the mismatched units between the model and the residual image that form the restored (cleaned) image when combining observations obtained with significantly different baselines \citep[see][]{Czekala_2021}. To test this hypothesis, we subtracted the residual image from our cleaned image to measure the remaining flux. The resulting image still exhibited a flux excess (without residuals) implying that the "JvM" effect could not explain the discrepancy.
\\ \\
We inspected the J2253+1608 blazar used as an amplitude calibrator in Cycle 6. In Figure \ref{calibrator} we can see the fluxes reported in the ALMA archive during 2019 March-May and July-September. Our high and low resolution data were taken on August 21 and 23 and April 27, respectively. The red square represents the J2253+1608 flux used to calibrate the low resolution data. The flux value is close to the one reported by the archive a day later and we considered that it yields a reliable calibration. The green triangles represent the J2253+1608 flux considered to calibrate each MS of the high resolution data. In this case, there were no measurements of the J2253+1608 flux close to the date of the observations, so it had to be interpolated between August 15 and 31. This interpolation is not robust as the calibrator is a variable source, with a flux that can rise and fall very quickly within intervals of days, as seen between the first part of the August light curve. Furthermore, one of the MS was calibrated with a flux that was not close to the interpolation, making the calibration for the high resolution data not reliable. Therefore, we scaled the fluxes in the combined [C$\textsc{ii}$] and continuum data using a scaling factor that allowed us to reach the level of flux seen in the low resolution data.

\end{document}